# From the firm- to the application-level: A new typology of AI use in organisations

Fulvio Castellacci
University of Oslo

November 2026

## Abstract

Artificial intelligence is applied in organisations in many different ways. However, the firm-level literature on AI lacks a conceptual framework able to describe that variety. Empirical studies typically examine one dimension of AI use at a time. This paper argues that the difficulty follows from the unit of analysis: a firm may hold many AI applications at once, each with its own novelty, sourcing mode, objective and technology type. These dimensions are defined at the level of the application rather than of the firm. We therefore propose shifting the unit of analysis from the firm to the individual AI application, and develop a typology combining these four conceptual dimensions. The new framework locates any application in a common space, and it points out in particular two types that the prevailing adoption-innovation distinction neglects: *co-invention*, novel uses of technology the firm did not develop, and *internalisation*, in-house building of AI that could have been bought instead. Documented use cases illustrate the framework, which is then used to discuss implications for firms' strategies and the diffusion of AI in economic systems.

# 1. Introduction

Research on artificial intelligence and its economic effects has grown faster over the past five years than almost any other area of economics and management, and public discussion of it has grown faster still. Recent research on the economic effects of AI has concentrated at the firm level, using a variety of data sources, such as ICT surveys, business and executive surveys, patents, online job postings, web-scraped firm data and corporate disclosures. This literature provides some important results and stylised facts. AI adoption is low, concentrated in large, young and already digitalised firms (Acemoglu et al., 2022; McElheran et al., 2024; Bonney et al., 2026; Pulito et al., 2026); and estimated productivity effects are on average positive but still limited in magnitude (Calvino and Fontanelli, 2026).

In spite of the important advances, a general difficulty faced by this recent research is that AI applications in firms and public organisations are multifaceted. AI can be used for virtually every activity and working operation that organisations carry out. Current research finds it difficult to cope with this complexity and variety of applications. This new literature often defines its central terms loosely. Firms are described as adopting AI, using it, investing in it, implementing it or innovating with it, often interchangeably, and studies rarely state which of these their data actually record. Specifically, what distinguishes a firm that adopts AI from a firm that innovates with it, and,what lies between the two?

To illustrate this problem, consider this example (World Economic Forum, 2026). In 2025, the company KPMG implemented the AI assistant that SAP sells alongside its enterprise software and applied it to the firm's own record of past system migrations, consisting of two hundred thousand pages of internal documentation, together with terabytes of curated material from previous projects. Using this huge background information, the resulting assistant knows how the firm has solved these software migration problems before, which customisations tend to break, which conversions are safe, and what went wrong previously. When the AI assistant was implemented, coding errors fell by half, and consultants reported to save around ninety minutes a day. In this AI application, no new model was trained by the firm, and no new method devised. The capability was bought by an external provider, and the same assistant is on sale to KPMG's competitors. What was unique was the company's archive, and the work lay in assessing which of three million documents were worth providing the AI assistant with. The resulting capability is nonetheless one no competitor can simply buy, because what makes it work is not the assistant but the archive behind it. In short, it is hard to say whether the AI application in this example should be classified as simply adoption of generative AI, or process/organisational innovation, or something in between the two.

The lack of conceptual clarity and precise definitions of different types of AI in organisations is also reflected in the empirical patterns and results that the firm-level literature reports. Current research presents a great deal of measurement dispersion. Estimates of firm-level AI adoption vary widely. This partly depends on how adoption is defined and measured. Bick et al. (2026) is an example of this: asking United States firms whether they use AI to produce goods or services returns 7%, and asking about use in any business function returns 17%. Relatedly, estimates of productivity effects due to AI use do also vary substantially in different recent studies.

The starting point of this paper is that current research on the economic effects of AI finds it hard to cope with the complexity and multidimensional nature of AI use in organisations because it has developed largely outside innovation studies, and it has therefore inherited neither the vocabulary nor the distinctions that innovation research built for precisely this problem over the preceding forty years. Innovation studies have for several decades used conceptual categories and empirical knowledge that may now be useful to provide more conceptual clarity to the firm-level literature on AI adoption, innovation and their economic effects. Specifically, the Oslo Manual (OECD/Eurostat, 2018) distinguished innovations that are new to the firm from those that are new to the market. Pavitt (1984) developed a taxonomy of innovation that distinguished different sources and modes of innovation. The classical distinction between product, process and organisational innovation is directly relevant to what firms are now doing with AI, but it is largely absent from the empirical literature on AI adoption. Hence, the underlying motivation of this paper is to use established conceptual categories from innovation studies to contribute to the literature on AI and its economic effects in firms and public organisations.

The more specific claim we make is that the main issue hampering current firm-level research on AI is that different works examine different dimensions of AI use at a time. Some studies focus on the source of AI (internally developed or externally purchased); others point out the type of AI technology; a few record what the AI is for. Most existing studies, though, do not consider together and combine these different dimensions. The paper argues that to understand and analyse the complex variety of AI applications in organisations it is necessary to consider several dimensions at once rather than one at a time.

The main reason why extant research typically focuses on one dimension and neglects others is the level of analysis at which the literature works, and the available data and indicators that it focuses on: the firm-level. When the unit of observation is the firm, a firm may hold many AI applications simultaneously (e.g. a purchased assistant, a new deployment of a commodity model, an in-house replication of something standard), each with a different sourcing mode, a different degree of novelty, a different technology and a different purpose. Conceptually, these dimensions are actually defined at the level of the application rather than of the firm; therefore, firm-level measures of AI use should be built up from application-level information rather than recorded directly. The implication is that the firm-level literature finds it hard to conceptually define and empirically measure the complex variety of AI applications. The present paper argues therefore that current research on AI adoption and innovation in organisations should shift its unit of analysis from the firm to the individual AI application. An application is a combination of a technology, a business function and a purpose, held within an organisation.

Specifically, the article develops a new typology of AI use in organisations by focusing on the four key conceptual dimensions briefly noted above. The first is the *novelty* of the application, i.e. whether an application is new only to the firm (technology adoption), or new to its market (innovation). The distinction is standard and it comes from the Oslo Manual. The second is the *sourcing mode*, i.e. whether the firm develops the AI system in-house, or acquires it externally. This dimension is also standard in innovation studies (e.g. Pavitt, 1984), and Babina (2026) has recently used it in the AI field. The third dimension is the *type* of AI technology that is being applied – analytical, perceptual or generative – which also defines the complementary assets that each type requires (Teece, 1986). The fourth relevant dimension refers to the *objectives* of AI use, i.e. whether the firm is applying AI to achieve more efficiency and cost reductions, higher sales and market shares, or to search for new knowledge and innovations. In turn, these different objectives will arguably have different economic effects, and it is therefore crucial to distinguish them. These dimensions are not new; they are well-known conceptual pillars in innovation studies. However, by considering the four dimensions together, the paper develops a new typology that is able to describe and classify a large variety of possible different AI applications in organisations.

The new typology points out four main results. First, adoption and innovation are endpoints of a space, not a binary dichotomy as often assumed in the literature. In particular, the typology uncovers two important types of AI applications that are neither pure adoption nor innovation. One is *co-invention*, in which new AI applications are built from technology the firm did not create (Bresnahan and Greenstein, 1996; Bresnahan, 2024; Bresnahan et al., 2025). The other is *internalisation*, in which a firm builds in-house AI technology that others have already developed.

Second, the typology shows what type of innovation (e.g. product, process, organisational or marketing) a given AI application may lead to. Existing AI research often neglects this aspect: only four of the 44 recent studies reviewed in this paper record the objective of AI use at all, and the type of innovation that this leads to. This addition is important. Much of what AI produces at present does arguably lead to process and organisational innovation, and a focus and measurement tradition built around product innovation systematically under-records it.

Third, each dimension implies an adoption cost, which increases with the complexity of the technology: the costs of organisational adjustment rise with novelty; the required complementary assets are larger for in-house development than external purchase; and the difficulty of showing that an application has paid off is higher as the objective moves from efficiency to market and from market to search. Hence, it is reasonable to expect that AI applications do not spread evenly across all types of the new typology, but they will rather concentrate in specific directions, each with different expected economic impacts. To examine this aspect, we provide an empirical illustration of the typology using three large sources of documented AI applications: the United States Federal AI use case inventory, a near-census of 3,611 reported applications; the United Kingdom Algorithmic Transparency Recording Standard, a register of records published by public bodies; and the World Economic Forum's 2025 MINDS cohort, which documents a large number of applications in private firms. Using these inventories, we identify real examples of different types of AI applications for each of the cells of our typology, and we analyse the distribution and frequency of these different types. In this sense, the typology is not only a device for classifying AI systems: by associating each position with the complementary assets it requires and the effect it can be expected to produce, it yields expectations about which types of AI application should be more frequent, and those that may be expected to have a stronger economic impact (Doty and Glick, 1994).

Fourth, when we discuss implications of the typology for the firm level, we point out a richer account of organisational AI profiles than the prevailing distinction between adopters and innovators allows. A firm often holds a diverse portfolio of applications, and its profile is given by the set of types in this portfolio. The framework further indicates what it takes for an organisation to acquire a new type over time, and the two routes out of pure adoption differ sharply in their requirements. The co-invention trajectory calls for organisational capacity and domain knowledge, whereas the internalization trajectory calls more for specialised human capital and ICT infrastructure.

Taken together, these results point to one general contribution. Firm-level measures of AI should best be seen as *compositions*: they are averages taken over portfolios of heterogeneous applications, and the main puzzles in this literature (the wide dispersion of adoption and productivity estimates, and the persistence of the adoption-innovation binary) are consistent with this aggregation. The proposed solution is to shift the focus to a different unit of observation, because the relevant conceptual dimensions are not properties that a firm can possess at all. One may ask whether an application is new to its market, and whether a particular system was built or bought; neither question is well posed of an organisation as a whole. Shifting to the application level is therefore what makes these categories well defined, and it is what allows co-invention and internalisation, two types that no data source currently in use is able to express, to be studied.

The paper is organised as follows. Section 2 sets out the four conceptual dimensions. Section 3 reviews the recent firm-level literature on AI, pointing out what this literature does and does not distinguish. Section 4 develops the new typology in four steps: novelty crossed with sourcing mode; the resulting types crossed with objectives; the same types crossed with technologies and the assets they require; and the combined typology that the three previous steps together define. Section 5 provides an empirical illustration of the typology using recent inventories of AI applications. Section 6 discusses implications for the study of AI use in organisations, and the dynamic trajectories of these. Section 7 concludes, and draws implications for national economies and for policy.

## 2. The conceptual dimensions

We start by pointing out the four conceptual dimensions that will be used to assess current research and then to develop the new typology of AI use in organisations. These dimensions are core pillars in innovation studies, and they have been much used in the innovation literature to provide the foundations of a large number of empirical studies. Some of these dimensions have also been recently used in a few studies within the firm-level literature on AI and its economic effects, as we note below.

**Novelty of the application.** A widely used distinction in the innovation literature is the one between a technology or application that is new-to-firm *versus* new-to-market, as noted in the Oslo Manual (OECD/Eurostat, 2018), which has organised innovation measurement for three decades. In the AI literature, this distinction is applied by using the terms AI *adoption* (i.e. a firm adopts and puts into use an AI technology that was developed elsewhere) and AI *innovation* (i.e. a firm develops a new AI technology). This simple distinction is important, and it is at the core of the firm-level literature on AI, as we will show in the next section.[1]

However, this binary distinction misses an important intermediate category that is neither adoption nor AI innovation: *co-invention*. The concept has its roots in the literature on general-purpose technologies, which observes that a technology of broad applicability generates value only where users adapt it to their own circumstances (Bresnahan and Trajtenberg, 1995). Bresnahan and Greenstein (1996) named this activity and documented it for business computing: turning a general capability into a working application required substantial inventive effort by the adopting organisation, which market-supplied tools could not easily reduce because it depended on knowledge of the user's own processes, sector and data. Hence, external advanced knowledge is usable only by an organisation that has built the internal absorptive capacity to recognise and adapt it (Cohen and Levinthal, 1990). Bresnahan and Greenstein (1996) further showed that co-invention costs rise with the complexity and idiosyncrasy of the application, which is why the same technology diffuses rapidly in some settings and only slowly in others.

[1] Our use of these terms departs from the Oslo Manual in one respect: in the Manual a business process that is new to the firm counts as an innovation, whereas we use the term innovation for applications that are *new to the market*, because our question concerns where the economic effect of an AI application originates rather than how innovation should be counted in survey measurement.

The concept has recently been discussed in the context of AI. Bresnahan (2024) argues that the bottlenecks determining whether AI becomes a general-purpose technology lie in co-invention rather than in the technology itself; Bresnahan, Greenstein and Yin (2025) distinguish incremental from novel co-invention, showing the latter to be rare but the most valuable; and Lang et al. (2026) study co-invention as the process by which general AI capabilities are translated into sector-specific applications. Co-invention is thus a third category standing between adoption and innovation: the firm invents nothing at the level of the technology, yet it does something for which there is no precedent at the level of the application. Recognising this category requires one modification to the way the novelty question is usually asked. The Oslo Manual poses it of the innovation itself, that is, of a product or a business process that is new to the firm or new to its market; we pose it of the application, i.e. a combination of technology, business function and sector, which is what allows a firm using an entirely standard technology to be doing something genuinely new to its market. Because novelty is assessed relative to that combination, and relative to what other competitors have already done, the category does implicitly refer to a given point in time: an application that is new to its market at one point becomes an established use once competitors imitate it, so each AI application must be classified as of the date in which it is first introduced.

**Sourcing mode.** The second dimension focuses on the sourcing mode of technology. From the point of view of a company that wants to implement a new technology, the distinction is between a new technology that has been internally developed by the company that implements it, and those that have been developed by other companies and purchased by the company in question. This distinction has a long tradition in innovation studies, and it is a key pillar in Pavitt (1984) taxonomy of innovation in firms. The complementarity between internal development and external acquisition has been studied by Cassiman and Veugelers (2006), who find that internal R&D and external knowledge acquisition tend to reinforce rather than substitute for one another.

In recent works on AI, this distinction has been used by Babina (2026), in her review of data sources and indicators used to study AI. It is also used in OECD (2026); and it is measured directly in the French ICT survey used by Calvino and Fontanelli (2026) and in the European sample of Hoffreumon et al. (2024).[2]

**Objectives of AI use.** The third dimension focuses on the objectives that a firm has when it decides to adopt or develop a new technology. We distinguish three distinct types of objectives: (1) *efficiency*: the new technology is meant to improve efficiency of the production process (e.g. through process or organisational innovation), thus reducing costs of labor, energy or other input factors; (2) *market*: the application is directed at the organisation's external users, and is expected to lead to higher sales in the case of a business firm, or to better service for citizens in the case of a public agency (e.g. through product, service or marketing innovation);[3] (3) *search*: the new technology is meant to increase the firm's capability to search the technological space, and hence increase the productivity of R&D and lead to new innovations. This is what Cockburn et al. (2019) define as "invention in the method of invention", (see also Bontadini et al., 2026, and Arenas Díaz et al. 2026).

The table below summarises these three categories, and points out the link between the objective that a firm has when using AI, the type of innovation that this may lead to (if any), and the type of economic effects that this may have for the company. It is interesting to note that this third conceptual dimension can be empirically measured by means of some existing survey questions (e.g. in the American Business Trends and Outlook Survey, and the European ICT surveys). However, this important aspect has seldom been considered in the recent AI literature. As shown in the next section, in fact, only four of 44 firm-level studies take this aspect into explicit account.

[2] In practice, however, the implementation of an AI application may sometimes fall between these two poles, since an organisation may assemble an application from commodity or open-source components that it has neither bought as a working system nor developed itself. For simplicity, we will retain the binary distinction for the development of the conceptual framework and typology below, and we will explain in the empirical illustration in appendix C how such intermediate cases can be assigned.

[3] Since the paper seeks to study AI systems in general, regardless of if they are deployed in private firms or public organisations, we will use the category "market objectives" in a broad sense, i.e. including not only private market objectives (that business companies typically pursue), but also public services for citizens (carried out by public agencies). The key defining feature here is that the organisation uses AI in order to have a better management of, and interaction with, its external users – that are customers for private firms, and citizens for public organisations.

**Table 1: Relationships between objectives of AI use, innovation type and economic effects**

| Objective | Activities and tasks | Innovation type | Economic effects |
|---|---|---|---|
| **Efficiency** | Automation, cost reduction, and internal decision quality (risk scoring, diagnostics, forecasting). | Business process innovation (process and organisational). | Cost side; higher productivity. |
| **Market** | Customer engagement, personalisation, new channels, and customer-visible quality improvement. | Product innovation (goods or services). | Demand side; higher sales; greater take-up of public services |
| **Search** | AI used in innovation projects, whether formal R&D or less formalised exploration; and AI used to develop further AI. | Downstream innovation of any type. | Greater future innovation capacity. |

**Type of technology.** The fourth dimension that it is important to consider is the type of technology that we are looking at. A large literature in innovation studies has convincingly shown that opportunities, constraints and results of innovative activities are typically technology-specific, and it is therefore crucial to consider the characteristics of a given technological trajectory when studying its patterns, determinants and effects. Specifically, AI represents a broad set of diverse technologies, which have different characteristics and that are currently following distinct technological trajectories (Pulito et al., 2026).

An important point is also that different types of technologies pose different requirements to firms in terms of the amount and type of organisational resources that they require for successful implementation, so-called *complementary assets* (Teece, 1986). Extant research points out three types of complementary assets: (1) specialised human capital (data scientists, machine-learning engineers, and the technical staff able to build, integrate and maintain AI systems); (2) ICT infrastructure (data, servers, cloud capacity, connectivity, sensors, and the enterprise systems through which business information is captured and stored); (3) organisational capacity (i.e. the managerial ability to redesign workflows, absorb disruption to established routines, and sustain a project past the pilot stage). The third of these differs from the first two. It is close to what the strategic management literature calls a dynamic capability, the ability to reconfigure the resource base rather than simply to operate it (Teece et al., 1997), whereas specialised human capital and ICT infrastructure are operating resources that must be maintained continuously once acquired.

Considering these aspects, we will distinguish between three types of AI technologies. (1) *Analytical AI*: they predict, classify and score from structured data. The scarce assets they require are ICT infrastructure (pipelines, storage, data quality) and specialised human capital able to prepare data and interpret output. Organisational demands are comparatively light, because analytical systems usually fit into existing decision processes rather than replacing them. (2) *Perceptual AI*: these applications interpret images, video, sound and sensor streams. The scarce assets that firms need to adopt and apply these technologies are typically physical: cameras, sensors, edge computing, installation, and the engineering human capital to integrate them into a working operation. Data-science skills matter less than they do in the analytical AI, and physical capital matters more. (3) *Generative AI*: they produce language, code, images and other content. This type of technology is easier to implement in terms of pure adoption, but to develop new applications and innovations it arguably requires more than the other types. What it does require throughout is organisational capacity, because generative systems reshape how work is done.[4]

## 3. The firm-level literature

This section presents a short and targeted review of the recent firm-level literature on AI use and effects in organisations. It reviews studies whose unit of analysis is the firm, and that focus on organisational use of AI technologies and/or its economic effects at the firm-level. Using these criteria to identify relevant studies, we have collected a total of 44 papers, all published in the last five years or less. Appendix A provides a summary and overview of these 44 studies, coding each for the unit of analysis, the data used, and whether it considers, among other things, the four conceptual dimensions that were outlined in the previous section. This literature review is focused and targeted: it does not seek to provide a broad and comprehensive discussion of this literature, but it rather intends to highlight whether and how the four conceptual dimensions are considered in this literature. The underlying purpose is therefore to point out clearly the knowledge gap in extant research, and prepare the ground for the development of the new typology in the subsequent sections. Table 2 provides a summary overview of the reviewed literature.

Regarding the first conceptual dimension, 35 of the 44 studies examine the adoption of AI, and 9 examine its invention. No study examines the intermediate cases, i.e. firms that put commodity technology to an unprecedented use (co-invention). No data source and indicator records explicitly whether an AI application is new to the firm alone or new to its market, and no study reports results split on that basis. An exception is the taxonomy of Calvino et al. (2024) that covers both AI innovation and AI use, but it does so at sector rather than firm level, which is why it is not coded here. The co-invention category is therefore under-studied, and hard to measure given available data and indicators at the firm-level.

[4] The three types of AI technology noted here are in line with other recent works studying this dimension. It is a classification focusing on what the AI system does with data, and the complementary assets it requires: structured data and statistical capability; sensors, labelled data and physical integration; unstructured content with interface and oversight capability. It may be argued that the recent diffusion of agentic AI may represent a separate category that should be added to these three. However, we think that AI agency should not be regarded as a distinct technology but rather a different deployment mode of existing technologies, i.e. autonomy over multi-step action and tool use, which can in principle apply to both analytical, perceptual and generative AI technologies. Furthermore, the diffusion of agentic AI is still at a very early stage and not yet well documented and studied. For these reasons, we prefer not to consider it further as a separate category in the conceptual framework. In the empirical illustration presented later in the paper (section 5 and appendix C), when a real deployment case is described as an agent, it is assigned to the technology type of the model that does the underlying work rather than to a category of its own.

Second, the sourcing mode is only partly present in the literature. 17 of the 44 studies do implicitly indicate the sourcing mode of AI applications. However, in 11 of them this is not explicitly studied as a conceptual dimension, but fixed due to the type of data and indicator used in the empirical analysis. Specifically, patent-based studies observe only firms that develop AI, because a firm that merely buys does not appear in the data: this holds for Bacchiocchi et al. (2026), Dernis et al. (2021), Giczy et al. (2022), Hötte et al. (2023), Igna and Venturini (2023), Lang et al. (2026), Marioni et al. (2024), and Rishabh and Shreeti (2026). The same holds for trademark-based studies, such as Castaldi et al. (2026), since only firms that register a new offering enter the data. By contrast, studies of a single vendor's product observe only firms that buy, such as Chatterji et al. (2026) and Chen and Stratton (2026). Only six studies in the literature consider together and compare firms using different sourcing modes: OECD (2026), Calvino and Fontanelli (2023; 2026), Fontanelli et al. (2025), Hoffreumon et al. (2024) and Kazakova et al. (2020). Where comparison is made, it is informative and useful: Calvino and Fontanelli (2026) find that the productivity returns of developers survive controls for selection while those of buyers do not, while Hoffreumon et al. (2024) find that the two modes behave as complements rather than substitutes, so that firms that buy also build.

Regarding the third dimension, objectives of AI use is the aspect that has been most neglected in this literature. Only four studies of the 44 papers record what the AI is for, its intended objectives for companies, and only one study (Aghion et al., 2025) considers together and compares among different objectives. Acemoglu et al. (2022) record six adoption motivations, Bonney et al. (2024) the common uses reported by adopters, and Aghion et al. (2025) seven types of use estimated separately. Only Arenas Díaz et al. (2026) consider the third objective, search, i.e. treating AI as a method of invention rather than a means of producing or selling. The lack of information and attention on this conceptual dimension is important, because, as noted in the previous section, firms' objectives and strategies are closely linked to the types of AI innovation that companies develop (process, organisational, product, etc.) and hence the economic impacts that AI may have for firms.

The fourth dimension, type of technology, is sometimes considered, but never analysed as a key factor explaining differences in complementary assets and firm performance. 13 studies of the 44 in our sample record what kind of AI is at issue, but only four of those compare across types:

OECD (2026) sets generative against non-generative, Bencivelli et al. (2025) generative against predictive, Pulito et al. (2026) core AI against generative AI, and Allen (2026) reports LLM adoption separately from AI adoption as a whole. The Danish evidence (Pulito et al., 2026) shows why the distinction is important: adoption is “highly technology-specific”, with science-trained workforces predicting core AI and university-educated non-science workforces predicting generative AI; and with machine learning and natural language processing diffusing across business functions while other technologies stay confined to particular domains.

Besides looking at the four dimensions in isolation, it is also important to ask whether the literature has studied these dimensions together. The answer is negative. 18 of the 44 studies focus on one dimension and specify none of the remaining three dimensions. The few studies that consider different dimensions always focus on sourcing mode and type of technology. Sourcing mode never appears alongside objectives, and type of technology never appears alongside objectives, in any study in the set. The only work that compares across two dimensions is OECD (2026). In summary, an important conclusion of this targeted literature review is that the multidimensional character of an AI application – represented by its degree of novelty, sourcing decision, type of technology and objectives – has not so far been the object of empirical analysis. It is important to clarify that the main reason for this is arguably not the lack of scholarly interest in these dimensions, but rather the fact that the data sources and indicators used for these empirical analyses often provide a hard constraint to the type of question and dimensions that it is possible to investigate empirically. Babina (2026) has recently presented a comprehensive overview of all different data sources and indicators used in this literature. It is interesting to point out here whether and how each of these data sources makes it possible to measure the four conceptual dimensions highlighted in the present paper. Appendix B illustrates this. The table in Appendix B shows two things. First, most of the data sources used in the literature do not make it possible to properly take into account and measure the conceptual dimensions noted here. Second, the two data sources that are more useful in this respect, and that are more frequently used in the literature – i.e. official ICT surveys and patent data – provide complementary information. The former focuses on firms’ adoption of AI technologies, whereas the latter focus on development of new AI innovations. Neither of them, though, enables to consider the important space in-between adoption and innovation, such as co-inventions and new AI applications that lead to innovations that contribute to the diffusion of AI as a general-purpose technology.

**Table 2: Number of studies in the AI firm-level literature that take into account the four conceptual dimensions of our framework**

| Dimension | Studies considering the dimension | Studies comparing across categories within the dimension | Notes |
|---|---|---|---|
| **Novelty of the application** | 44 | 0 | 35 examine adoption, 9 examine innovation; no study records whether an application is new to the firm or new to its market; none considers co-invention |
| **Sourcing mode** | 17 | 6 | In most of the 17, the sourcing mode is implicitly fixed by the type of data and indicators used |
| **Objectives of AI use** | 4 | 1 | Aghion et al. (2025) estimate effects separately for seven types of use |
| **Type of technology** | 13 | 4 | Generative *vs* non-generative, generative *vs* predictive, core *vs* generative |

# 4. The new typology

We build up the new framework in four subsequent steps. Each of the first three steps produces a typology that considers some of the conceptual dimensions, but neglects others; finally, the fourth step considers all four dimensions, and sets the three typologies together into a combined typology of AI use in organisations. Table 3 presents the logic of exposition followed in this section, and it points out the question that each step focuses on. In this exercise, as noted in the previous sections, the unit of analysis is the individual AI *application*, not the firm. Type names denote kinds of activity and not firms, and a single firm will typically hold applications in several combinations at once (an aspect that we will discuss further in section 6). The typology developed in this section is not meant to be merely a scheme for sorting cases. Since each AI type that will be identified determines the complementary assets that an application of that kind requires and its possible impacts in terms of innovation and economic performance, the typology is meant to be analytically useful rather than purely descriptive (on this distinction, see Doty and Glick, 1994).

**Table 3: The typology development in four steps**

| Step | Typology | Conceptual dimensions | Number of AI application types | What it answers |
|---|---|---|---|---|
| Step 1 | Typology 1 | Novelty of application; sourcing mode | 4 types | Does an AI application represent adoption, innovation, or something else? |
| Step 2 | Typology 2 | Novelty of application; sourcing mode; objectives of AI use | 12 types | What does an AI application produce: which innovation type, through which economic channel? |
| Step 3 | Typology 3 | Novelty of application; sourcing mode; type of AI technology | 12 types | What does an AI application require: which complementary assets, at what level? |
| Step 4 | Combined typology | All four dimensions | 36 types | All three questions at once |

## 4.1 Typology 1: Novelty and sourcing mode

In the first step, we only consider the two first conceptual dimensions noted in sec. 2: the novelty of the application and the sourcing mode. The first dimension looks whether the application is new only to the firm (AI adoption) or new to the market (AI innovation). The second dimension focuses on whether the firm develops the AI system in-house or sources it externally, hence whether the firm acts as a *producer* of the technology, or merely a user of it.[5] When we consider together these two dimensions, we identify four distinct types of AI applications. Two of these are standard and widely used in the literature. *AI adoption* is external sourcing applied to an established use. *AI innovation* is in-house development applied to an unprecedented use. However, the other two types are usually neglected by extant empirical research. *Co-invention* is external sourcing applied to an unprecedented use. As noted above, this is a category that has been conceptually introduced by Bresnahan and colleagues, but it is usually neglected in firm-level empirical studies. Finally, we point out a fourth type that has not been previously considered in the literature, and that we call *internalisation*. By internalisation, we mean in-house development of a technology already developed and commercialised by others. The reason why organisations may decide to develop internally a technology that already exists may be due to two different reasons. One is that they do not want to buy external technologies for a given internal use, e.g. due to regulation, privacy issues and other ethical concerns (something that may be relevant for many public organisations, for instance). The other is that they want to build up internal capabilities in AI development, given the crucial strategic importance that AI has for the organisation.

[5] It follows that when one organisation builds an AI system and another puts it to use, these should be considered two distinct applications rather than one. The application is classified from the point of view of the organisation that holds it, so the same system appears as AI innovation for its developer and as adoption for the organisation that buys it.

**Table 4: Typology 1. Question: Does an AI application represent adoption, innovation, or something else?**

| | APPLICATION NEW TO THE FIRM ONLY | APPLICATION NEW TO THE MARKET |
|---|---|---|
| **AI IS DEVELOPED INTERNALLY** | **INTERNALISATION** In-house replication of an established capability | **AI INNOVATION** Frontier technology, unprecedented application |
| **AI IS BOUGHT EXTERNALLY** | **ADOPTION** Commodity technology, established application | **CO-INVENTION** Commodity technology, new application |

## 4.2 Typology 2: Adding objectives and types of innovation

In the second step, we extend the first typology by adding the third conceptual dimension: the objectives that the organisation has when it introduces a given AI application. As noted previously, objectives are important because they are closely related to the types of innovation that the AI application entails, and the economic impacts that it will have for the firm. Specifically, an efficiency objective leads to process and organisational innovation; a market objective produces product or marketing innovation; a search objective raises the firm's capacity to innovate downstream. When we add this third dimension to typology 1, we obtain a second typology, that extends the first, and that points out 12 different types of AI applications. Table 5 shows this second typology. The various examples that we note below are purely illustrative (in sec. 5, we will instead provide real-world cases of AI applications).

**Adoption – Efficiency:** This is arguably the most frequent type of AI application in organisations: commodity tools applied to existing internal work. *Example:* A law firm subscribes to a commercial contract-review tool that flags non-standard clauses (standard use, no adaptation).
**Adoption – Market:** Standard customer-facing AI bought off the shelf. What distinguishes this combination is that the AI is visible to customers yet identical across competitors, so it confers no advantage: firms adopt defensively, to avoid falling behind, rather than to get ahead. *Example:* A retailer subscribes to a standard personalisation and recommendation service (new to the retailer; standard in its market).

**Adoption – Search:** Standard AI platforms used within innovation projects. Conceptually important, since it is the clearest case where the purpose is innovation while the AI use itself is not new. *Example:* A pharmaceutical firm runs a well-established commercial screening platform in its discovery pipeline, similarly to what its competitors do.

**Co-invention – Efficiency:** Bought AI applied to a new internal process, leading to process innovation and productivity gains. *Example:* A salmon farming firm applies off-the-shelf computer vision to underwater video to count sea lice and estimate biomass (standard models; no precedent in aquaculture; the core production process is materially changed).

**Co-invention – Market:** Bought AI assembled into a new customer-facing offering, leading to product or service innovation and increases in demand. *Example:* An insurance broker assembles commercial language models into an automated claims-triage service and offers it to clients as a new product line (no new model was trained; a new service is introduced).

**Co-invention – Search:** A novel application of bought AI whose output is further innovation. *Example*. A construction-materials producer applies commodity language models to fifty years of unstructured failure reports and site logs, using them as a hypothesis-generation instrument to identify candidate mechanisms of material degradation, which then feed a physical testing programme (the technology is bought; the application is new; the output is downstream innovation).

**Internalisation – Efficiency:** This is the core archetype of internalisation, in which AI is built in-house because data cannot leave the organisation. *Example:* A public agency builds an internal chatbot over its own document archive rather than licensing an equivalent product, because data sensitivity and vendor data-residency rules forbid external sourcing.

**Internalisation – Market:** Building a proprietary customer-facing system where commercial equivalents exist. *Example*: A large grocery retailer builds its own recommendation and personalisation engine rather than licensing one, because customer transaction data is a strategic asset it will not expose to a vendor (recommendation engines are entirely standard; the decision is about data control, not novelty).

**Internalisation – Search:** In-house R&D or simulation tools built where commercial equivalents exist, again for governance or intellectual-property reasons rather than novelty. *Example:* A pharmaceutical firm builds its own compound-screening and simulation pipeline, duplicating capability available commercially, because the compound library and the screening criteria are core intellectual property that cannot be exposed to a third-party platform.

**AI Innovation – Efficiency:** This happens not when AI is applied to the firm's business processes, but when it is applied to the firm's own AI infrastructure. For instance, a novel method (e.g. quantisation, model compression, caching, workload scheduling) that reduces the compute, memory or energy cost of systems the firm already runs. *Example*: A large telecommunications operator develops an unpublished model-compression method that substantially cuts the inference energy cost of its network-optimisation systems, and keeps it internal rather than licensing it.

**AI Innovation – Market:** This type denotes AI-native products and ventures, that is common for many firms studied in the entrepreneurship literature on AI. *Example*: A maritime software firm trains its own foundation model on vessel sensor time-series, where no such model exists, and builds a predictive-maintenance service on it (both the technology and the application are new).

**AI Innovation – Search:** This is a recursive combination: frontier methods used in innovation search, including AI used to develop further AI (e.g. model-assisted engineering, synthetic data generation, architecture search). *Example:* A drug-discovery firm develops a new machine-learning method for predicting molecular binding that exists nowhere else, and uses it to generate new therapeutics.

**Table 5: Typology 2. Question: What does an AI application produce: which innovation type, through which economic channel?**

| | EFFICIENCY | MARKET | SEARCH |
|---|---|---|---|
| **ADOPTION** | Commodity tools for internal work. New-to-firm process change; no new-to-market novelty. | Standard customer-facing AI. New-to-firm product innovation. | Standard platforms used in R&D. |
| **CO-INVENTION** | Novel internal application of bought AI. Process innovation. | Novel customer-facing service built on bought AI. Service innovation. | Novel use of bought AI in innovation search. |
| **INTERNALISATION** | Built in-house for governance reasons. Process innovation. | Proprietary customer-facing systems. Common for large firms only. | In-house R&D tools where commercial equivalents exist. |
| **AI INNOVATION** | Novel AI methods that reduce the cost of the firm's own AI. | AI-native products and ventures. | Frontier AI methods in R&D; and/or AI used to build AI. |

### 4.3 Typology 3: Adding technology types and required complementary assets

The third step is analogous to the second step, as it extends the first typology by adding another conceptual dimension. The additional dimension considered in this step is the type of AI technology that characterises a given application (i.e. dimension four in our framework). As discussed in section 2, the type of technology is conceptually closely related to the type and extent of complementary assets that a given AI application requires. In particular, analytical AI requires first and foremost ICT infrastructure and specialised human capital; perceptual AI applications necessitate physical capital assets and the engineering human capital to use and integrate these. As for generative AI, these are often easier to adopt and use, but to develop new applications and innovations they do require substantial organisational and managerial capacity. When we add this fourth dimension to typology 1, we obtain a third typology, that extends the first, and that points out 12 new different types of AI applications (which differ from those noted in the previous step). Table 6 illustrates this third typology. The logic of the grid is that position on the vertical axis determines the *level* of complementary assets required, which rises as one moves from simple adoption towards more complex AI innovation. On the other hand, position on the horizontal axis determines the *kind* of complementary assets that the organisation needs. A firm may be well endowed in one kind and poorly in another, so the two axes together determine not simply whether an application is demanding, but whether a particular firm can undertake it.

**Analytical AI - Adoption.** The technology is bought; the assets around it are not. *Example:* A retailer subscribes to a commercial demand-forecasting service and supplies it with its own sales and stock data (the model is standard; the work and the cost lie in getting clean data into it).

**Analytical AI - Co-invention.** A standard method is applied on a dataset that did not previously exist. *Example:* A district-heating utility applies standard gradient-boosting methods to newly instrumented pipe-network sensor data in order to predict failures before they occur (the methods are off the shelf; the data did not exist before, and no operator had built a failure-prediction application on it).

**Analytical AI - Internalisation.** This type requires the full in-house stack (specialists, pipelines, monitoring and retraining). *Example:* A bank develops its own credit-scoring model instead of licensing one, because supervisory requirements and the sensitivity of the training data prevent it leaving the bank's systems.

**Analytical AI - AI innovation.** It requires research-level human capital, proprietary data at scale, and substantial compute. *Example:* A transmission system operator develops a new forecasting architecture for grid balancing, trained on decades of its own network telemetry, because no existing method handles weather, storage and demand response at that resolution.

**Perceptual AI - Adoption.** Purchased systems require capital and physical integration, so the barrier is substantial even at the adoption level. *Example:* A food processor installs an off-the-shelf vision system on its packaging line to detect seal defects (a standard product used as intended; the expense is installation and line integration more than the software).
**Perceptual AI - Co-invention.** This type of application adds deep domain knowledge and substantial organisational capacity to the capital requirement, because a new operating routine must be built around the sensing. The complementary assets are physical, local and slow to replicate. *Example:* A salmon farming firm applies off-the-shelf computer vision to underwater video to count sea lice and estimate biomass (standard models; no precedent in aquaculture; the core production process is materially changed). This is the same example already noted under Co-invention – Efficiency, seen now from the technology side.
**Perceptual AI - Internalisation.** This case may arise for organisations whose imagery cannot leave their premises due to security, regulatory or related reasons. *Example:* A defence contractor builds an in-house model to detect and classify vehicles in aerial imagery, because the imagery is classified information and cannot be transmitted to or processed by an external provider (the method is standard; the data handling is not).
**Perceptual AI - AI innovation.** It requires research-level vision expertise together with a physical test environment, which is a combination few organisations hold. *Example:* A firm develops new network architectures and training regimes for object recognition in turbid water, training from scratch because standard vision models assume clear optical conditions and fail below a certain visibility threshold (the methods are new rather than fine-tuned, and they enable inspection tasks that were previously not possible).
**Generative AI - Adoption.** It usually does not demand data pipeline and specialists, which is precisely why adoption spread so quickly after 2022. The binding constraint is managerial: someone must redesign the workflow so that the tool is used, train staff, and carry the change. *Example:* A municipal administration rolls out a commercial generative assistant for routine correspondence and meeting minutes (a standard subscription used as intended; the difficulty is getting staff to change how they draft, not the tool).
**Generative AI - Co-invention.** This is the application type that has the lowest technical requirement and the highest organisational one. What it requires is a document archive, domain judgement, and the managerial capacity to turn an internal capability into a service. *Example:* A law firm builds a case-law summarisation service for a specific area of domestic regulatory practice using commodity language models over its own archive (first to market nationally, and imitable by any competitor in a short time span).

**Generative AI - Internalisation.** It requires infrastructure to host models securely and specialists to maintain them, plus data governance. *Example:* A hospital trust hosts an open-weights language model on its own infrastructure to draft summaries of clinical notes, because patient records may not be sent to an external provider.

**Generative AI - AI innovation.** This is arguably the most demanding application type in the whole framework: it necessitates large curated corpora, compute at a scale few organisations possess, and capable research staff. *Example:* A language-technology firm trains a foundation model for its national language from scratch. It counts as AI innovation: the firm trains the model itself rather than fine-tuning one supplied externally, and no adequate model for that language existed. Had a satisfactory model been available and rebuilt in-house for data-privacy reasons, the same effort would count as internalisation.

**Table 6: Typology 3. Question: What does an AI application require: which complementary assets, at what level?**

| | ANALYTICAL | PERCEPTUAL | GENERATIVE |
|---|---|---|---|
| **ADOPTION** | Data pipelines and analytical staff, even for a bought system. | Capital and physical integration; a significant barrier even off the shelf. | Low technical requirement. High organisational capacity. |
| **CO-INVENTION** | Proprietary data newly made observable. | Capital, engineering and deep domain knowledge. | Lowest technical and highest organisational demand in the grid. |
| **INTERNALISATION** | Full in-house stack. Gated by scale. | Only where imagery cannot leave the premises. | Secure hosting and governance. |
| **AI INNOVATION** | Research staff, proprietary data at scale, compute. | Research vision expertise plus a physical test environment. | Corpora, compute, research staff. |

### 4.4 Step 4: the combined typology

In the fourth and final step, we combine the previous three typologies together. Considering all four conceptual dimensions at once, we define a space of thirty-six AI use types. The typologies introduced in the previous sub-sections are best understood as projections of this combined space. Typology 2 is the projection onto the novelty-by-objective plane; Typology 3 the projection onto the novelty-by-AI technology plane. Each is a four-by-three view of the same underlying space, taken from a different angle: one showing what applications produce, the other what complementary assets they require. Figure 1 shows the combined typology space. Each plane is Typology 2, and there is one plane for each of the three technology types of Typology 3. Reading one AI use type across the three planes gives the novelty-by-technology view of Typology 3. Hence, each AI use type is a cell with three coordinates, and the shaded cell in figure 1 is simply an example (co-invention, for search objectives, using analytical AI technology).

Why is this combined typology useful? The typology makes it possible to position any given AI application that we may think of in a space constructed by means of four clear and well-established dimensions in innovation studies, from which its innovation type and its complementary asset requirements can be discussed. This is a comprehensive and useful conceptual framework that scholars and students of AI can use to navigate the complexity and multifaceted nature of AI use in organisations.

An important implication follows directly from the framework. Each of the dimensions carries a cost that arguably rises in one direction: organisational adjustment increases with the novelty of the application, complementary assets rise with in-house development, and the difficulty of demonstrating a return from AI investments increases from efficiency through market to search objectives. AI applications are therefore expected not to spread evenly across the 36 types, but concentrate towards the adoption of commodity technology for established, efficiency-oriented uses, and to be less common in the opposite and more demanding direction. Section 5 will illustrate this further.

**Figure 1: The combined typology (4 X 3 X 3 = 36 types of AI applications).**

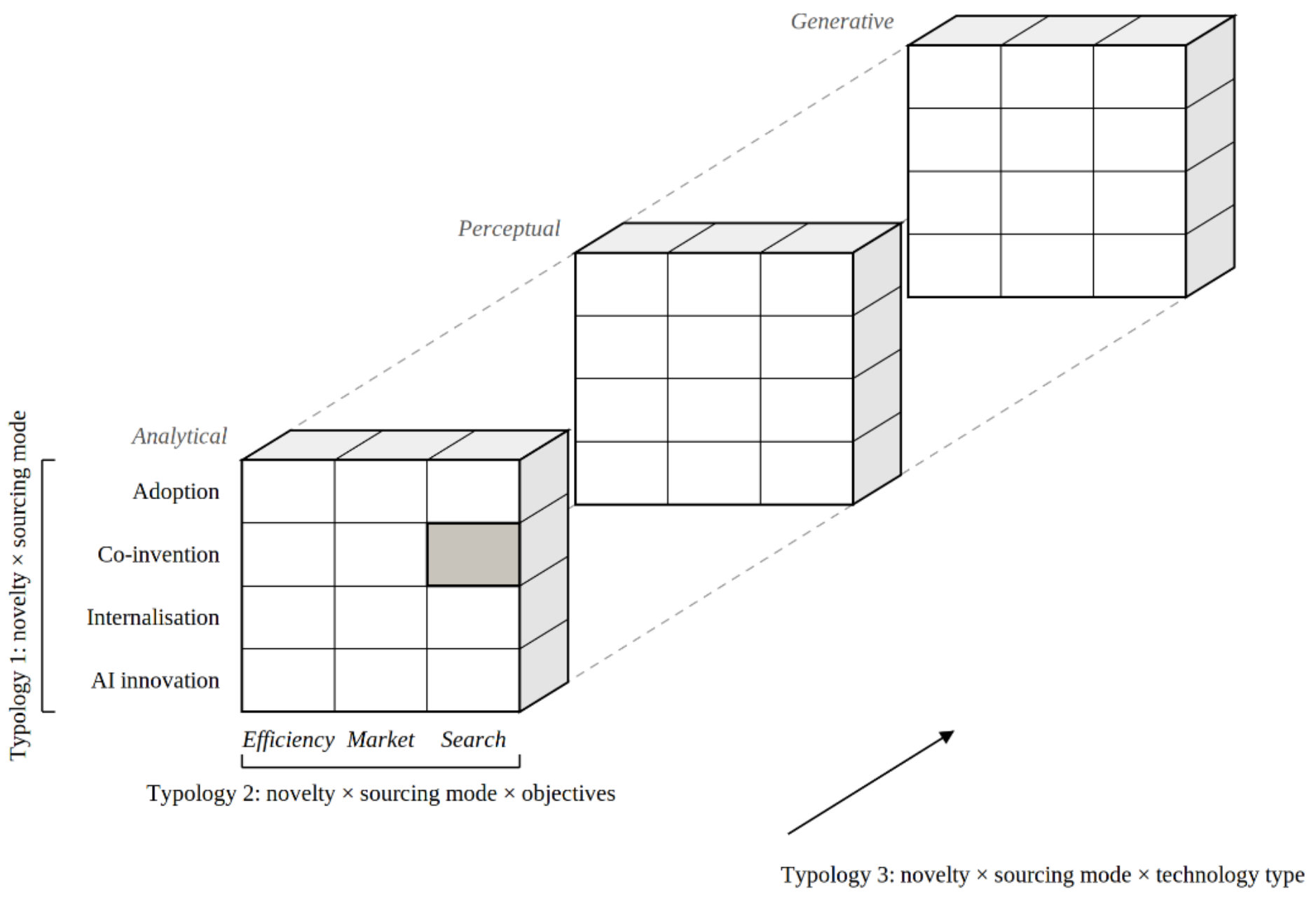

## 5. The new typology: An empirical illustration

This section presents an empirical illustration of this combined typology. The purpose is not to carry out an empirical test to show the validity of the typology. Rather, we seek to provide descriptive and anecdotal evidence to show that the typology is a useful tool to illustrate many different types of AI use in organisations. To do so, we use three original data sources: these are not standard statistical databases in tabular format, but inventories of real and recent AI use cases, in (more or less structured) text format.

**Inventory 1: The World Economic Forum 2025 MINDS cohort.** The report *Proof over Promise: Insights on Real-World AI Adoption from 2025 MINDS Organizations*, produced jointly by WEF and Accenture and released in January 2026 (World Economic Forum, 2026), documents named organisations selected for demonstrated and measured impact, spanning a large variety of AI use cases.[6] It draws on hundreds of cases across more than thirty countries and twenty industries, with thirty-two cohort entries in the geographical listing. The inventory is not meant to be a representative survey but simply an illustration of selected cases.

[6] This report is downloadable at: https://www.weforum.org/publications/proof-over-promise-insights-on-real-world-ai-adoption-from-2025-minds-organizations/.

**Inventory 2: The United States federal AI use case inventory.** This is a consolidated register published by the US Office of Management and Budget as a public repository.[7] Federal agencies are required to report whatever AI they use, so the register is close to a census of its population rather than a selection. The records for the year 2025 contain 3,611 applications.

**Inventory 3: The United Kingdom Algorithmic Transparency Recording Standard.** A mandated register of algorithmic tools used by central government departments and, increasingly, local authorities and health bodies.[8] It holds 141 published records as of August 2026. There is no downloadable dataset and the register is too small to support distributional statements, so it is used here purely as a source of well-documented individual cases.

What is interesting about these three data sources is that they contain rich and detailed textual information about a large number of different AI use cases in organisations, both private firms and public agencies. The description of each recorded case makes it possible to describe, or to infer, the four conceptual dimensions used in this paper. However, the coding procedure differs across the inventories. For the American federal inventory, it is rule-based and applied to the register's own structured fields: the topic field to the objective of use, the technology classification to the technology type, and the develop-or-acquire field to the sourcing mode. Of the 3,611 records, 1,140 carry enough information to be placed on these three dimensions. For the MINDS cohort and the United Kingdom register, where the information is narrative rather than structured, we used a large language model (Claude Opus 5) to read the descriptions and proposed a coding against rules fixed in advance by the author. The coding was then checked case by case against the source texts in a second step, and the discrepancies identified were examined and corrected by the author. Subsequently, the 33 MINDS records were re-coded again from the source independently of the original coding, which reproduced the distribution reported below. By doing so, we are able to position a large number of these real use cases in the various cells of the combined typology.

The result of this exercise is that we are able to classify real AI use cases drawn from the three inventories to the various cells of the combined typology. Table 7 shows the typology space and an example of a use case that we have identified for each cell. Appendix C provides a more detailed description of each case noted in this table. 15 cases are illustrated from the MINDS cohort, 17 from the American federal inventory and 4 from the UK register.

---

[7] The inventory is available on a public GitHub repository at https://github.com/ombegov/2025-Federal-Agency-AI-Use-Case-Inventory. The file that we have used is "2025_individually_reported_AI_use_cases".

[8] The inventory is available at: https://www.gov.uk/algorithmic-transparency-records. There is no bulk download and no application programming interface, so records must be opened individually.

Beyond finding a real case for each cell of our combined typology, we also carried out a simple descriptive analysis of all the cases reported in the three inventories (in addition to the 36 reported in table 7 and appendix C). An interesting pattern concerns two of the dimensions that these inventories allow us to cross-tabulate: the objectives of AI use and the type of AI technology. First, considering the WEF MINDS inventory, table 8 shows that efficiency objectives are much more common (54%) than market and search. As for technology type, analytical technology is much more frequent (57%) than the other two types (in spite of the recent rapid growth of gen-AI), and perceptual AI the least represented. Second, for the US federal inventory, table 9 points out a similar pattern. Efficiency objectives account for nearly 72% of coded use cases, while market objectives account for only 6%.[9] For technology type, perceptual applications are the least represented technology group (16%), whereas analytical and gen-AI are much more frequent.[10] The sourcing mode can also be observed directly in the American federal register (where it is compiled for 44% of the 3,611 records). Of these AI applications, 37% were purchased from a supplier, 36% were developed in-house, and 27% were developed with a combination of contracting and in-house resources. Since federal agencies are not suppliers of AI technology, in-house development in this population consists for the most part of building capabilities that are commercially available, which is what this paper defines as *internalisation*. This type therefore accounts for more than a third of the applications for which the sourcing mode is recorded. The share built in-house also differs markedly across technology types, from 52% of analytical applications to only 15% of perceptual ones, which is consistent with the patterns reported in table 6. The novelty of an application, by contrast, is not recorded as a field in any of the three inventories, so the frequency of co-invention cannot be established in the same way. On the whole, the observation that some types of AI applications are more frequent than others is important, and it may have implications for the study of firms, that we discuss in the next section.[11]

[9] As defined in sec. 2, the label "market objectives" is also meant to include public services for citizens, when the organisation is a public agency rather than a private business company.

[10] The two dimensions coded from the register's structured fields were assigned as follows. From the "Use Case Topic Area" field, efficiency was assigned to Administrative Functions, Information Technology, Law Enforcement, Health and Medical, Energy and the Environment, Human Resources, Cybersecurity, Procurement and Financial Management, and Emergency Management; market, which in this paper also includes services addressed to citizens, to Service Delivery, Government Benefits Processing, Transportation and International Affairs; and search to Science and to Other – Applied Mathematical Sciences. From the "AI Classification" field, analytical was assigned to classical and predictive machine learning and to reinforcement learning, perceptual to computer vision, and generative to generative AI. Records classified as natural language processing or as agentic AI do not correspond to a single technology type and were not coded on this dimension.

[11] A caveat concerning this empirical illustration relates to the nature of these data sources. Public sector registries (the US Federal and UK inventories) and corporate cohorts (the WEF MINDS), due to the nature of organisations they focus on, may exhibit compositional skews. For instance, public agencies face strict data-sovereignty mandates, security protocols, and public accountability rules that may bias these inventories toward internalisation and efficiency objectives. Similarly, firms cohorts more commonly highlight successful, high-impact deployments.

**Table 7: Empirical illustration of the combined typology: a real AI use case for each cell of the typology.**

| | EFFICIENCY OBJECTIVES | | | MARKET OBJECTIVES | | | SEARCH OBJECTIVES | | |
|---|---|---|---|---|---|---|---|---|---|
| | **Analytical** | **Perceptual** | **Generative** | **Analytical** | **Perceptual** | **Generative** | **Analytical** | **Perceptual** | **Generative** |
| **ADOPTION** | **1** Hypotension alerting device | **2** Colonoscopy lesion detection | **3** GitHub Copilot | **4** HMRC digital assistant | **5** One Login liveness check | **6** Public helpline bot | **7** Pangolin lineage classifier | **8** Map text extraction | **9** Claude Enterprise |
| **CO-INVENTION** | **10** Fujitsu: supply chain | **11** PepsiCo: 3D vision | **12** Genshukai–Fujitsu: hospital agents | **13** NICE: market forecasting | **14** Landing Med: cervical screening | **15** Ant Group: clinical platform | **16** UCSF–SandboxAQ: screening | **17** AI4Peat drainage mapping | **18** Sanofi–OAO: discovery agents |
| **INTERNALISATION** | **19** DWP fraud triage | **20** NASA queue counting | **21** SEC internal assistant | **22** Hearings triage | **23** USCIS: checking of applicant photos | **24** Public comment triage | **25** Plasmid classification | **26** Plover habitat mapping | **27** ATSDR tox chatbot |
| **AI INNOVATION** | **28** Hyundai–DEEPX: edge AI | **29** Siemens–EthonAI: inspection | **30** Cambridge Ind.: private LLM | **31** Horizon Power–TerraQuanta | **32** Saudi MoH–AmplifAI: thermal imaging | **33** Tech Mahindra: sovereign LLM | **34** Synopsys: RL chip design | **35** USGS: wave-runup segmentation | **36** NASA JPL: ROSA |

Note: Appendix C provides a more detailed description of each case noted in this table, and the source from which it is drawn.

**Table 8: Cross-tabulation of use cases in the MINDS inventory (N=33).**

| | Analytical | Perceptual | Generative | All |
|---|---|---|---|---|
| **Efficiency** | 30 | 12 | 12 | **54** |
| **Market** | 9 | 6 | 9 | **24** |
| **Search** | 18 | 0 | 4 | **22** |
| **All** | **57** | **18** | **25** | **100** |

**Table 9: Cross-tabulation of use cases in the US federal inventory (N=1140).**

| | Analytical | Perceptual | Generative | All |
|---|---|---|---|---|
| **Efficiency** | 26.9 | 13.4 | 31.5 | **71.8** |
| **Market** | 2.6 | 0.5 | 3.2 | **6.4** |
| **Search** | 17.0 | 2.5 | 2.2 | **21.8** |
| **All** | **46.6** | **16.5** | **36.9** | **100** |

## 6. Implications for the firm level

The unit of analysis of our conceptual framework and new typology is the individual AI application. Data used in current research, however, are typically collected at firm level, and the questions that matter for policy and for performance are therefore often asked of firms. This section discusses implications of our framework for the study of AI at the firm-level. Three clarifications should be made before this discussion. One is that a firm may acquire new application types and lose others over time; hence, when we refer to the organisation's *AI profile*, we define it as the set of types of AI use it holds *at a given time*. Second, when we discuss a firm profile, we do not refer merely to the number of AI applications it holds, but rather the significance that these have for the organisation's strategy and performance. Third, although the typology of applications uses four dimensions, the firm profiles discussed below are defined on the first two only. Novelty and sourcing are properties that an organisation either holds or lacks (based on the applications they have), and they can therefore be used to position firms in that space; objectives and technology type vary across the applications that a firm holds, and so they describe the composition of a given firm profile and the complementary assets it requires, as discussed in sec. 4.3, rather than its position.

## 6.1 Firm mixed profiles

A firm may hold a diverse portfolio of AI applications, each of which occupies a position in the typology space developed in section 4. Where one single type dominates, we define the profile as *pure*; where two or more types are present, we define it as *mixed*. Typology 1 can be used to define four pure firm profiles: the adopter, the co-inventor, the internaliser and the AI innovator firms. As shown in section 3, the firm-level literature typically studies only two of these, the adopter and the innovator, and neglects the other two.

What our framework adds here is that organisations may hold mixed AI profiles. That organisations use different types of AI systems rather than only one is not merely a conceptual possibility. For instance, drawing on survey data for more than 3,000 European firms, Hoffreumon et al. (2024) find that buying ready-made AI software and developing AI in house are complements rather than substitutes in most sectors, so that the organisations which build are typically also the organisations which buy. Further, in the French ICT survey, Calvino and Fontanelli (2026) find that nearly 15% of AI users both buy and develop AI systems, so that roughly one AI user in seven holds a mixed profile on the sourcing dimension. The reason for these previous findings in the literature is basically that buying and building draw on the same internal capacity: an organisation that develops AI is thereby better placed to recognise and adapt what it buys, i.e. it has the sufficient and required absorptive capacity, as noted in section 2.

Using the space and categories defined by typology 1 (see Table 4 in sec. 4.1), four positions define six possible pairwise combinations, i.e. six different mixed profiles. Figure 2 shows these six profiles. Each panel represents one (stylised type of) organization. The two shaded cells, outlined in bold, mark the positions where the organisation's applications predominantly belong to.[12]

**(1) Adoption plus co-invention.** The firm buys its AI externally, but it also puts some of what it buys to a new use for which there is no precedent in its market. This profile resembles the supplier-dominated innovator described by Pavitt's (1984) taxonomy: the firm invents nothing at the level of the technology, yet it does something new at the level of the application.

**(2) Adoption plus internalisation.** The firm applies established capability throughout. It buys most of it, but it also builds internally a few AI applications due to governance or regulatory reasons, e.g. because the data cannot leave the organisation (e.g. as when commercial marketing tools are combined with an in-house credit-scoring model).

[12] Profiles that span three or four positions are of course also possible; but for simplicity we disregard them here. In the discussion below, when we point out a given mixed firm profile, this is defined by predominance rather than by exclusivity, meaning that a firm is assigned to one of these profiles based on where the bulk of its AI activity is located, although the same firm may well hold other occasional and less important AI applications of other types.

**(3) Co-invention plus internalisation.** The firm finds novel uses for the AI technology it buys, and it also builds in house capability that it could have bought, due to governance or regulatory reasons. This type of organisation does neither ordinary AI adoption nor frontier development work.

**(4) Co-invention plus AI innovation.** Everything the firm does with AI is new to its market, and hence innovative. Some of it is assembled by applying commodity components to new uses, and some other is developed internally from scratch.

**(5) Internalisation plus AI innovation**. The firm develops its AI internally. Some of the AI systems it builds replicates capability that was available on the market; while another part represents development of new AI technologies.

**(6) Adoption plus AI innovation.** The firm buys its periphery off the shelf (adoption), while at the same time it also builds its core AI systems at the technology frontier. This type resembles the idea of an *AI-native venture*, i.e. a firm that has AI development as its core purpose and specialisation since its very establishment.

One implication of this analysis concerns firm size. Five of the six mixed profiles noted here include a position in which the firm develops its own AI. As set out in section 4.3, AI development requires specialised staff and infrastructure that are concentrated in larger organisations. By contrast, smaller firms are more likely to be either pure adopters, or adopters who also co-invent. Large firms are more likely to hold mixed profiles because they are able to do many things with AI across many functions. AI-native micro-firms are mixed for the opposite reason: they do very few things with AI, and those few lie at opposite ends of the space. Mixed profiles are therefore not simply more frequent the larger the firm. They can be present at both ends of the size distribution, and they are arguably least likely in the middle: a medium-sized firm may hold enough AI applications not to be pure, but it has neither the range of business functions of a large firm nor the sharp division between a bought periphery and a built core of an AI-native venture.

## 6.2 Dynamic firms' trajectories: What it takes to change profiles

The discussion above has described static firms' profiles, that is, what an organisation's AI use looks like at a given time. However, AI is a fast-diffusing technology, so it is also worth asking a related question: how these profiles change over time, and how organisations can scale up their AI use to increase its economic benefits. By definition, almost every organisation begins as a pure adopter. Buying AI initially does not require substantial assets beyond the subscription and the managerial attention needed to use it. This is why adoption of generative AI spread so quickly after 2022. The question is what comes after the first adoption stage. The account that follows in this and the next sub-section is conjectural: it sets out the possible trajectories and constraints that the framework leads us to expect, and the expectations we formulate are meant to be tested in future empirical research rather than being results established here.

Figure 3 points out two distinct trajectories. Both begin from an organisation that only buys AI externally (adopter). In the first route, the *co-invention trajectory*, the firm takes technology it has already bought and puts it to a new use for which there is no precedent. In the second, the *internalisation trajectory*, the organisation builds in-house a capability that it previously bought. Each step adds one position to the portfolio, so each turns an initial pure profile (adoption) into a mixed one. The two routes are not alternatives in principle, and a firm may take both. They are not, however, equally available to all firms.

The asymmetry between them lies in what each trajectory requires. Co-invention draws on organisational capacity and domain knowledge: the managerial work of redesigning a routine around a new technology, and enough understanding of the firm's own operations, data and sector to see a use that no competitor has tried before. As noted in section 2, neither can be bought, and neither depends on firm size. In generative AI, where the technical requirement is low, a small professional-services firm with a good archive may do this better than a large bureaucracy. By contrast, internalisation draws on the availability of specialised human capital and ICT infrastructure. These are largely fixed and indivisible, and they must be maintained rather than acquired once. As Hopf et al. (2023) observe from the study of 55 implementation projects, running a machine-learning system is not a one-time project but a continuous effort that requires specific in-house expertise. Therefore, in the terms of section 2, the first trajectory builds a capacity to reconfigure, while the second trajectory builds a function that has to be operated. This second trajectory often carries a threshold of size, given its requirements in terms of resources and infrastructures, reason why most small firms find it difficult to follow it.

What happens if an organisation seeks to move even further, and go from a scaling up stage towards an AI development and innovation stage to compete at the technology frontier? In principle, the second step on each of the two trajectories will extend what the firm already does. On the co-invention trajectory, the firm extends adaptation into applications that are new to the market and builds them itself. On the internalisation trajectory, it extends in-house development beyond replication, thus developing new to the market algorithms. Both trajectories would arrive at a profile that contains AI innovation as its latest step, which is the position requiring both acquisitions at once: an application that is new to its market and built in-house demands the domain knowledge to see the use and the engineering capacity to construct it. We would not expect a firm to reach this position in a single move from pure adoption. Of the two endpoints, internalisation plus AI innovation is likely to be the more common, because building at scale may produce occasional novelty on its own. On the other hand, co-invention plus AI innovation requires novelty to be sustained across an entire portfolio, against competitors who seek to imitate.
Four qualifications are important to understand the dynamic trajectories presented in figure 3. First, the various stages indicated in this model are not obligatory steps. A firm may stop at entry (adoption), or at scaling up, and most firms never reach the technology frontier, lacking the resources or the capabilities to go further. Absorptive capacity is cumulative: an organisation that only buys does not build the capacity that the first step requires, so pure adoption represents a stable strategy rather than only a cheap and convenient choice. Second, the shared timeline in figure 3 aligns the two trajectories stage by stage, but the model makes no claim that they necessarily proceed at the same pace. If anything, the internalisation trajectory is arguably the slower, since it requires investments and capability building before it can begin. Third, in figure 3, a position left unshaded at a later stage has not necessarily been abandoned by the firm. What changes along a trajectory is where the organisation's applications are predominantly located: an application that counts as adoption at the first stage becomes, in the second stage, ordinary technology use, and therefore not anymore a defining feature of the firm AI profile. Fourth, the third stage is labelled *technology frontier* rather than market leadership, to indicate that a position in this space is not necessarily meant to imply a given level of performance. These trajectories are conceptual propositions, and their implications in terms of economic performance is an interesting topic for future empirical research.
One further case does not fit these two stylised firm trajectories, and it is therefore not shown in Figure 3. In the mixed profile called *adoption plus AI innovation*, which figure 2 labelled as the *AI-native venture*, where the firm buys its periphery and builds its core at the frontier. An established adopter cannot reach this profile by the steps just described, since it would have to cross the in-house threshold and reach the frontier at the same time, while continuing to buy ordinary software for everything else. AI-native firms do not move towards the frontier over time by following the stages noted in figure 3; they just begin at the frontier, because they are established from the very beginning having AI development as their core specialisation and strategic core. Hence, what explains this type of mixed profile is not a dynamic trajectory and evolution over time, but rather the very founding of the firm.

### 6.3 What it takes to maintain profiles

For any organisation, moving to a new and more advanced AI profile is one thing; however, holding it over time is another. Two forces work against the maintenance of a given firm profile over time. The first is imitation. An AI application counts as co-invention because it is new to its market at a given point in time. But once a competitor has replicated it, which may sometimes take a relatively short time, the application is no longer unprecedented and is reclassified as ordinary adoption. The same force acting on a frontier application returns it to internalisation: the firm still built it, but others now have it too.

The second force that challenges firms' AI profiles over time is commoditisation. An AI application counts as internalisation because the firm built what it could otherwise have bought. When the supply side catches up and an equivalent application becomes purchasable at a fraction of the cost, the in-house version is no longer worth maintaining, and the firm eventually buys it instead. The application is then reclassified as adoption. Similarly, a frontier application affected in the same way returns to co-invention.

Both forces, imitation and commoditisation, push in the same direction and act continuously: a firm's AI profile tends to return back to adoption. Over time, a firm that does nothing cannot maintain its competitive position: it drifts back towards pure adoption. This gives a dynamic reading of the static picture documented throughout the paper. Adoption is not only the stage where most firms begin; it is also where the system tends, and a firm is a co-inventor only for as long as it keeps producing co-inventions.

The two forces do not fall evenly on different types of firms and organisations. Imitation is likely to fall hardest on small firms, because the applications within their reach are the cheap ones, generative co-invention in particular, and cheap applications are the easiest to copy. The natural defence is to consolidate, building the capability in-house so that what a competitor can copy is the idea rather than the system. But that defence means taking the first step on the internalisation trajectory, which is the step that small firms find hardest in the first place. Their AI innovation is therefore structurally temporary in a way that the same activity in a large firm is not. Commoditisation, by contrast, affects mostly firms that build AI: it releases resources previously tied up in maintaining something that can now simply be bought.

**Figure 2: The six mixed firm profiles (represented in the typology 1 space).**

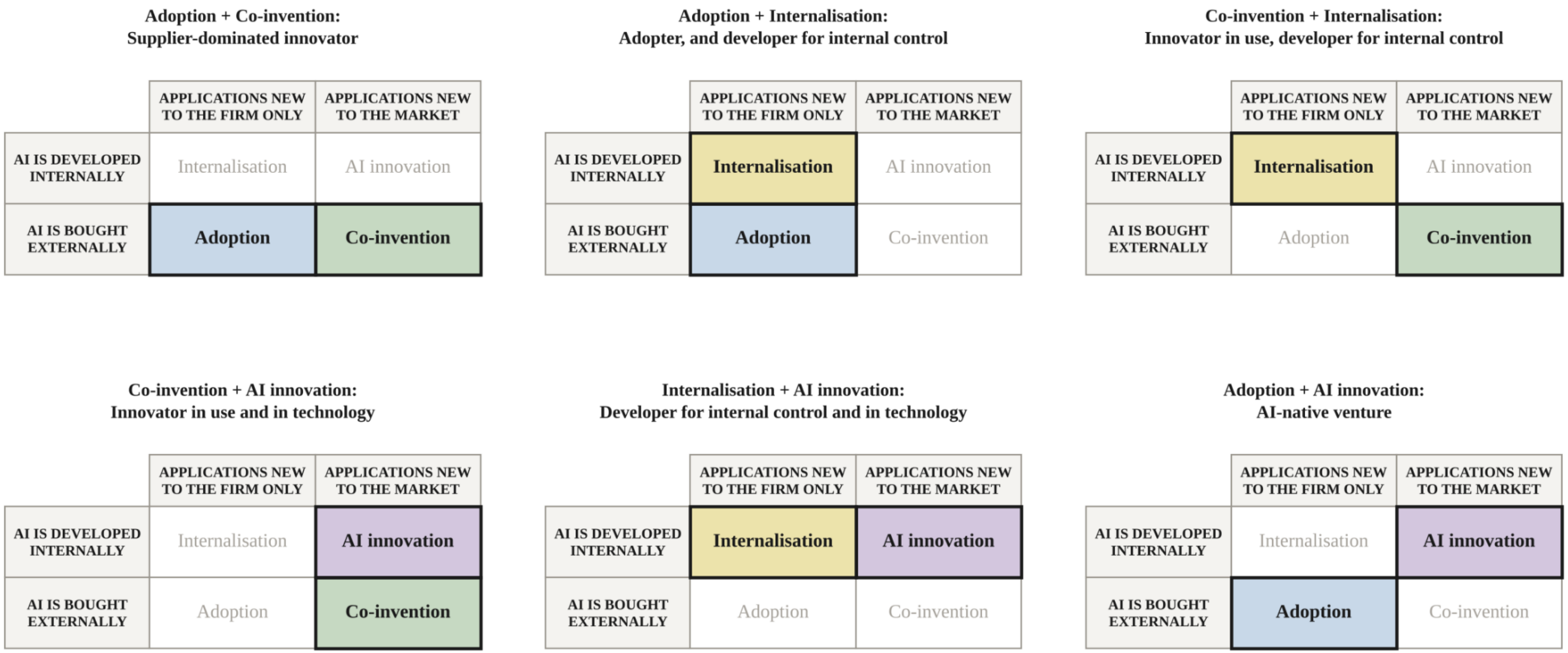

**Figure 3: The dynamics of AI use in organisations: Two distinct trajectories**

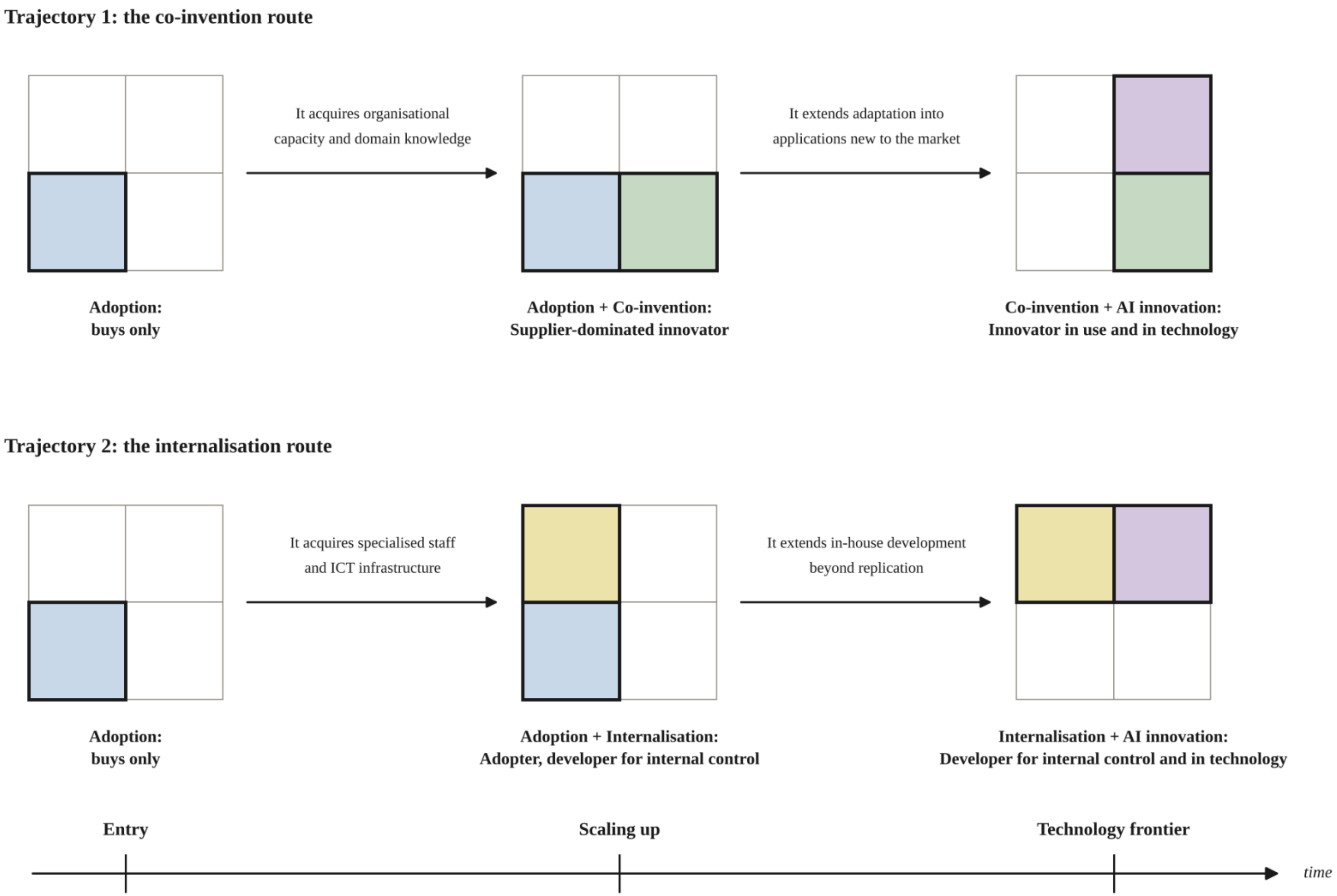

## 7. Conclusions

Research on the economic effects of artificial intelligence has expanded rapidly, but it still finds it difficult to conceptualise, analyse and measure the rich and multidimensional nature of AI uses in organisations. Empirically, the recent literature presents a range of different results. Estimates of firm-level adoption rates as well as reported productivity effects differ substantially. This paper has argued that these conceptual difficulties and empirical disagreements are partly the consequence of the level at which AI is being studied.

The argument rests on a simple observation. A firm does typically not use AI in one way only. It often holds a portfolio of applications, and these differ in how novel they are, in whether they were bought or built, in what they are meant to achieve, and in the kind of technology they rest on. A firm-level measure of AI is therefore an average taken over a distribution that the researcher never observes. This paper has proposed instead that the unit of analysis should be the individual AI application. Focusing on that level, we developed a new conceptual framework that is based on four dimensions long established in innovation studies: the novelty of the application, the sourcing mode, the objective pursued and the type of AI technology. Crossing these four dimensions defines a space of thirty-six positions, built up through three intermediate typologies, each of which answers a different question about an application: what it is, what it produces, and what it requires.

This conceptual framework points out four main findings. First, adoption and innovation are the endpoints of a space rather than the two sides of a dichotomy, and other important types of AI lie between them. Second, different types of AI applications lead to distinct types of innovation (product, process, organisational, or marketing), which in turn may have different economic impacts in terms of productivity, economic growth and labor demand. It is therefore paramount to go beyond the general concept of AI innovation, and specify further what innovation type a given AI application may lead to. Third, AI applications are not spread evenly across that space: using three inventories of real and well documented use cases, we noted a marked concentration in efficiency objectives. Fourth, once the typology is taken back to the firm-level, firms are found to hold mixed profiles and distinct dynamic trajectories that the prevailing binary conceptual categories in extant research cannot describe.

Our framework opens up new questions for the firm-level empirical literature on AI and its economic effects. If firm-level AI variables are compositions of diverse AI applications that each firm has, then the dispersion of adoption estimates present in the literature may be a consequence of the fact that available data and indicators at the firm-level average over different parts of a firm's AI portfolio. A firm's applications can fall into any of the four types of Typology 1, and most firms of any size may hold several types at once. But when a survey asks the firm one question and records one answer, four values are combined into one, and the only distinction that survives is whether the firm used AI or developed it. Dispersion in adoption estimates has actually several well-documented sources besides composition (e.g. how adoption is defined and what threshold of use counts, the sampling frame and whether estimates are weighted by firm size, the year of the survey, and whether the respondent is the firm, an executive or a worker). Composition should therefore be read as one possible candidate explanation among others. Different measures of productivity effects reported at the firm-level may have the same cause. A statistic counts a firm as an AI user whether it has bought a single assistant for drafting emails or rebuilt its core production process around a system of its own. Most users are of the first type, and for them no strong effect on performance should be expected. A measure that counts firms, instead of weighing what each firm actually does with AI, will therefore show widespread use but also little measured effect. What the paper suggests is that aggregation may be sufficient to generate these patterns, and that no data sources that are currently used in the firm-level literature on AI can rule it out, since none of them records composition of AI applications for a given company.

The typology is what makes this claim operational. It gives any AI application a specific position with four coordinates, from which the type of innovation it represents and the complementary assets it requires can both be derived. The new typology provides a framework to understand further the multifaceted nature of AI in organisations, and it supplies a vocabulary of major AI use types, to which results obtained from different data sources can be compared.

This conceptual framework and typology may be relevant for two user groups in particular. For researchers studying AI use and its economic effects, it supplies a unit of observation and a set of coordinates that can provide a more detailed understanding of AI adoption and how it differs among organisations. For statistical offices and survey designers, the framework specifies what would have to be added to existing instruments at the firm-level.

An implication goes beyond the individual organisation. According to our framework, a national economy's realised effect from AI may be thought of as the sum, across the positions of the typology, of the number of applications in each type multiplied by the effect that position produces. Position on the novelty dimension determines whether there is an effect at all, since adoption by construction reproduces what is already standard; the mix of objectives then determines what kind of effect it is, whether higher productivity, higher demand, or greater innovative capacity in the future. The aggregate outcome therefore depends on the composition of a country's AI applications rather than on its level of adoption. Two economies with identical adoption rates may differ substantially in how their applications are distributed, and should differ correspondingly in what they obtain from them. This may offer an account of cross-country differences that adoption rates alone cannot provide, and it may be one reason why the international evidence on AI and productivity has so far been so difficult to interpret.

The framework also connects this literature to growth theory. In models of general-purpose technologies, the aggregate effect of a new technology is delayed because complementary components must be invented before AI can be used productively (Brynjolfsson, Rock and Syverson, 2021). That co-invention effort is central to the mechanism, but it enters such models as an abstract quantity with no observable counterpart. The share of an economy's applications located in the co-invention row is relevant: an economy whose applications are mainly adoption is in the phase during which limited aggregate effect can yet be visible, whereas an economy with a substantial co-invention population is in the phase during which the technology is being converted into productivity. This idea is related to what Bresnahan and Greenstein (1996) showed for earlier general-purpose technology: the pace of co-invention is governed by the organisational complexity of the applications involved rather than by the rate of technical progress upstream. However, as noted in sec. 6, composition is a flow rather than a stock. Because imitation and commoditisation continually return applications to adoption, a favourable composition observed at a given time provides no guarantee of persistence: it has to be maintained and renovated over time. Industrial structure matters here. If smaller firms are exposed to imitation, their co-invention is often only temporary. Economies whose innovative activity is concentrated in smaller firms should therefore show less persistent co-invention shares than economies with a larger-firm structure, even where the two appear similar in a cross-section. This idea may provide a link between national industrial structure (and average firm size) and a country's AI outcomes and economic performance. Whether this holds is an empirical question for future research.

For policy, our conceptual framework suggests that supporting adoption rates alone may be a less effective policy objective in advanced economies, where the returns to commodity technology applied to established uses may be more modest, precisely because such applications demand little and are available to every competitor. A more important policy objective should instead be to shift composition, moving AI applications from adoption towards co-invention. That target implies a different policy approach. If the binding constraint on co-invention is organisational capacity and domain knowledge rather than access to technology, then subsidising access to tools, which is what most current policy programs do, does not address the binding constraint, while managerial capability, workflow redesign and the diffusion of organisational practice are the aspects that matter the most.

## Declaration of generative AI and AI-assisted technologies

As noted in section 5, the large language model Claude Opus 5 was used to classify the AI use cases reported in the three data inventories (the US federal AI use case inventory, the UK Algorithmic Transparency Recording Standard, and the WEF's 2025 MINDS cohort) into the various cells of the typology. The same LLM was used to write the plotting code that produced figures 1 to 3 based on the author's design; the figures are not AI-generated images. The LLM was also used to check language editing. The author reviewed and edited all output and takes full responsibility for the content of this paper.

# Appendix A. An overview of the firm-level literature on AI use in organisations.

| Reference | Unit of analysis | Country and period | Data sources to measure AI | Outcome variable | Main result | Novelty of AI application | Sourcing mode | Type of AI technology | Objectives of AI use |
|---|---|---|---|---|---|---|---|---|---|
| **Panel A. Firm-level studies of AI adoption** | | | | | | | | | |
| Acemoglu et al. (2022) | Firm | United States, 2016–2018 (2019 ABS) | Annual Business Survey technology module | Adoption; labour productivity; labour share; employment | Adoption low and concentrated in large and young firms; adopters have higher productivity and lower labour shares | Adoption | No | No: AI against four other automation technologies | Six adoption motivations |
| Allen (2026) | Firm and worker (three surveys compared) | United States, up to 2025 | Comparison of three public surveys: BTOS firms, Real-Time Population Survey workers, Survey of Business Uncertainty executives | Adoption levels and trends | About 18% of firms, 41% of workers and 78% of the labour force | Adoption | No | Generative (LLM) against all AI | No |
| Bacchiocchi et al. (2026) | Firm | Italy, 2012–2019 | Firm accounts matched to patent records; AI innovation identified from AI patent filings | AI innovation | Firms facing higher cash-flow volatility are significantly more likely to innovate in AI | Innovation | Develops (by construction: AI patents) | No: AI patents against more mature technology classes | No |
| Lang, Lien & Timmermans (2026) | Venture (new firms) | Norway, 2016 onward | Manual classification of AI ventures identified from a commercial directory and the trademark register | Entry trajectory; regional knowledge environment; relocation | Technology-first and domain-first co-invention trajectories are present in the same regions | Innovation | Develops (by construction; firms that only use AI tools are excluded) | No | No |
| OECD (2026) | Firm | Multiple countries, microdata 2023 | Multiple microdata sources combined | Competition; concentration; markups; venture funding | Non-generative adoption is not associated with rising market power; AI patenting concentration correlates with sales concentration and faster markup growth | Adoption | Users against developers | Generative against non-generative | No |
| Rishabh & Shreeti (2026) | Firm and region | Global, up to 2025 | Firm-level data on AI producers, mapped by location and position in the AI supply chain | Geographic distribution of AI production | AI firms cluster in a small number of locations | Innovation | Develops (by construction: AI producers) | No | No |
| Arroyabe et al. (2024) | Firm (SMEs) | European Union, 2020 | Flash Eurobarometer 486; self-reported adoption of AI technologies by SMEs | Adoption | Digital capabilities drive adoption | Adoption | No | No | No |
| Bencivelli et al. (2025) | Firm | Italy, 2020 and 2024 | Bank of Italy business surveys; self-reported adoption of cloud and AI | Adoption; expected employment | AI adoption is associated with size, exports and innovation capacity | Adoption | No | Generative against predictive | No |
| Bonney et al. (2026) | Firm | United States, Nov 2025–Jan 2026 | Business Trends and Outlook Survey AI supplement | Adoption; employment; firm performance | 18% of firms use AI in at least one function; 57% of adopters use it in three or fewer | Adoption | No | No | No: business functions, but not objectives |
| Bonney et al. (2024) | Firm | United States, Sep 2023–Feb 2024 | Business Trends and Outlook Survey; bi-weekly self-reported AI use for business purposes | Adoption; expected adoption | AI use is higher in large and young firms, with a non-monotonic size relationship | Adoption | No | No | Common uses reported: marketing automation; virtual agents; |

| Reference | Unit of analysis | Country and period | Data sources to measure AI | Outcome variable | Main result | Novelty of AI application | Sourcing mode | Type of AI technology | Objectives of AI use |
|---|---|---|---|---|---|---|---|---|---|
| | | | | | | | | | data and text analytics |
| Calvino et al. (2026) | Firm | 15 OECD countries, 2019–2023 | Harmonised code applied to national ICT surveys; adoption of five advanced digital technologies including AI | Adoption; productivity | Larger firms adopt more, and human and technological capital explain much of the adopters' productivity premium | Adoption | No | No: AI against four other advanced digital technologies | No |
| Castaldi et al. (2026) | Firm | Italy, up to 2023 | AI-related trademarks identified from goods and services descriptions, compared with AI patents | AI innovation | AI trademarks capture the commercialisation of AI in new goods and services, complementing patents and covering all sectors | Innovation | Develops (by construction: AI trademarks) | No: AI trademarks treated as one class | No: goods and services classes, not objectives |
| Dernis et al. (2021) | Firm | International, patents and trademarks, Up to 2019 | OECD micro-data lab; AI-related patents and trademarks matched to company-level data | AI innovations, goods and services | AI-related IP activity concentrates in a small number of large firms and countries | Innovation | Develops (by construction: AI patents) | No: AI-related IP treated as one class | No |
| Giczy et al. (2022) | Patent (linked to assignee firms) | United States, up to 2020 | Machine-learning classifiers applied to 13.2 million patent documents across eight AI component technologies | AI innovation | A validated dataset identifying AI in patents | Innovation | Develops (by construction: AI patents) | Eight technologies, including NLP; AI hardware; machine learning; computer vision | No |
| Hötte et al. (2023) | Patent | United States, 1990–2019 | Patents granted: four alternative AI patent classification approaches | Concentration and ownership of AI invention | The four AI classifications range from 3% to 17% of all US patents | Innovation | Develops (by construction: AI patents) | No | No |
| Udomsaph (2026) | Firm | 11 MENAAP economies, 2022–2025 | Keyword text analysis of open-ended innovation responses in World Bank Enterprise Surveys | Adoption | Innovation alone is insufficient for AI adoption; digital readiness and global integration are important complementary factors | Adoption | No | No | No |
| Yotzov et al. (2026) | Firm | United States, United Kingdom, Germany and Australia, 2025-2026 | Parallel questions fielded by four central-bank and university teams to almost 6,000 senior executives | Adoption; productivity; employment | 69% of firms use some AI; more than 90% of managers report no impact on productivity or employment | Adoption | No | No | No |
| Calvino & Fontanelli (2023) | Firm | 11 countries, 2018–2021 | Harmonised statistical code applied to official national ICT surveys | Adoption; productivity | AI use concentrated in ICT and professional services and in large firms; productivity premia largely explained by complementary assets | Adoption | Develops own algorithms (reported for a subset of firms) | No | No |
| Calvino & Fontanelli (2025) | Firm | France, 2020 and 2022 | French ICT survey | Adoption; technology specialisation | 6% of firms use AI; most specialise in one technology for one function, while larger firms adopt several technologies for different purposes | Adoption | No | Seven technologies, including text mining; image recognition; NLG; autonomous movement | No: business functions, but not objectives |
| Calvino & Fontanelli (2026) | Firm | France, 2018 | French ICT survey matched to registers | Adoption; productivity | Buyers have higher ICT endowment; developers are larger and younger | Adoption | Buys against develops | Predictive AI only (period characterisation, not a variable) | No |

| Reference | Unit of analysis | Country and period | Data sources to measure AI | Outcome variable | Main result | Novelty of AI application | Sourcing mode | Type of AI technology | Objectives of AI use |
|---|---|---|---|---|---|---|---|---|---|
| Hoffreumon et al. (2024) | Firm | European Union, Jan–Mar 2020 | European Commission enterprise survey on AI (Ipsos) | Adoption; sourcing strategy | Buying and building are complements rather than substitutes | Adoption | Buys against develops | Ten application categories, including NLP; computer vision; anomaly detection; creative generation | No: application categories, but not objectives |
| Li & Liu (2026) | Firm | Canada, 2019 and 2021 | Survey of Digital Technology and Internet Use linked to administrative business microdata | Adoption; productivity | Complementary capabilities predict adoption | Adoption | No | No | No |
| Lo Turco & Sterlacchini (2026) | Firm | France, 2018 and 2020 | French ICT survey; first-time AI adoption in 2020 with antecedents measured in 2018 | Adoption | Data security systems, ICT training and website-based collection of customer information predict first-time adoption | Adoption | No | No | No: business functions, but not objectives |
| McElheran et al. (2024) | Firm | United States, 2018 | 2018 ABS; self-reported use of five AI-related technologies | Adoption; diffusion patterns | Fewer than 6% of firms use AI; AI use clusters with cloud and other emerging technologies | Adoption | No | Five technologies: machine learning; machine vision; NLP; voice recognition; automated-guided vehicles | No |
| Pulito, Pytlikova, Schroeder & Lodefalk (2026) | Firm, linked to employer–employee registers | Denmark, 2023 and 2024 | Nationally representative Danish firm surveys linked to administrative registers | Adoption; breadth of deployment | Adoption is highly technology-specific; science-trained workforces predict core AI, non-science graduates predict generative AI | Adoption | No | Core AI against generative AI against other advanced technologies | No: deployment breadth across functions, but not objectives |
| Baslandze et al. (2026) | Firm | United States, 2025–2026 | Survey of corporate executives; self-reported AI investment and perceived productivity effects | Adoption; productivity; employment | More than 50% of firms have invested; productivity gains reflect increases through innovation and demand channels | Adoption | No | No | No |
| Bick et al. (2026) | Firm and worker | US and Europe, 2025–2026 | Combined worker and firm surveys across countries | Adoption; time savings; industry growth | Large adoption gaps between the US and Europe, partly explained by worker demographics, firm composition and management practices | Adoption | No | No | No |
| Calvino et al. (2022) | Firm | United Kingdom, 2020–2021 | Three big-data sources combined: intellectual property records, website text and online job postings, matched to firm accounts | Adoption | AI adopters concentrate in ICT and professional services and in the South of England; adopters are larger and more productive | Adoption | No | No | No |
| Chatterji et al. (2026) | Firm | Global, up to 2026 | ChatGPT Enterprise accounts, linked to worker roles, task classifications and public company accounts | Adoption; task composition | Adoption concentrates among larger and more R&D-intensive firms | Adoption | Buys (by construction: a single vendor's product) | Generative | No: message-level task classes, but not objectives |
| Dahlke et al. (2024) | Firm | Germany, Austria and Switzerland, 2019–2021 | Transformer language model applied to text from over 1.1 million websites | Adoption; diffusion | AI adoption follows epidemic diffusion operating through co-location, network embeddedness and social capital | Adoption | No | No | No |

| Reference | Unit of analysis | Country and period | Data sources to measure AI | Outcome variable | Main result | Novelty of AI application | Sourcing mode | Type of AI technology | Objectives of AI use |
|---|---|---|---|---|---|---|---|---|---|
| Hunt et al. (2024) | Commuting zone (from firm vacancies) | United States, 2007–2019 | AI publications merged with Burning Glass vacancy text | AI employment growth | Vacancies requiring AI skills grow more slowly farther from pre-2007 period | Adoption | No | No | No |
| Kazakova et al. (2020) | Firm | 30 European countries, Jan–Mar 2020 | European Commission enterprise survey on AI | Adoption; sourcing strategy; barriers | In-house building is less frequent, and it concentrates in large and technical firms | Adoption | Builds in house; customises; buys; AI as a service | Ten technologies, including NLP; computer vision; anomaly detection; autonomous machines | No |
| Yu et al. (2026) | Firm | United States, 2016–2025 | S&P 500 | Adoption; profitability; productivity | 11% of S&P 500 firms had AI deeply integrated in business processes, and 10% used it in production or service delivery | Adoption | No | No: generative AI and deep learning treated together | No |
| **Panel B. Firm-level studies of the effects of AI** | | | | | | | | | |
| Aghion et al. (2025) | Firm | France, 2018–2020 | Insee Technologies in Business survey, matched to employer–employee and balance-sheet data | Employment; sales | AI adoption is associated with higher firm employment and sales; effects are positive for ICT security and slightly negative for administrative processes | Adoption | No | No | Seven use types, estimated separately |
| Chen & Stratton (2026) | Firm | Global, 2021–2025 | Proprietary records of 200 million work events; firm-level adoption timing of GitHub Copilot and Cursor | Productivity; task composition; employment | Moderate productivity gains in code activity and task completion without quality loss | Adoption | Buys (by construction: vendor tools) | Generative | No |
| McElheran et al. (2025) | Establishment | United States 2017 and 2021 | Census collection on AI-related technologies with industrial applications | Productivity; profitability; inventory; robot investment; labour | Causal evidence of J-curve returns: short-run losses in older establishments precede longer-run gains | Adoption | No | No | No |
| Aldasoro et al. (2026) | Firm | EU and US, 2019-2024 | EIB Investment Survey matched to ORBIS | Labour productivity; employment; wages | AI adoption raises labour productivity by about 4% through capital deepening, with gains concentrated in large firms | Adoption | No | No | No |
| Arenas Díaz et al. (2026) | Firm | Spain, 2018 and 2022 | Encuesta sobre Estrategias Empresariales | R&D and technological change | AI is positively associated with both R&D and embodied technological change, and it triggers later innovation investment | Adoption | No | No | AI as a method of invention; R&D and embodied technological change |
| Fontanelli et al. (2025) | Firm | France, 2019 | French ICT survey matched to registers | Volatility of productivity growth | AI use raises the volatility of productivity growth; the effect is concentrated among buyers | Adoption | Buys against develops | Predictive AI only | No |
| Lu, Phillips & Yang (2024) | Firm and industry | China, 2010–2020 | Firm exposure to cloud infrastructure and AI, matched to entry, exit, M&A and financing outcomes | Industry dynamics; concentration; financing | Cloud lowers concentration; AI raises concentration | Adoption | No | No: AI against cloud computing | No |
| Marioni, Rincon-Aznar & Venturini (2024) | Firm | 15 EU countries, 2011–2019 | Patenting in AI fields, matched to firm accounts | Multi-factor productivity; | Firms that innovate in AI experience a productivity boost after developing the technology, | Innovation | Develops (by construction: AI patents) | No: AI patent fields treated together | No |

| Reference | Unit of analysis | Country and period | Data sources to measure AI | Outcome variable | Main result | Novelty of AI application | Sourcing mode | Type of AI technology | Objectives of AI use |
|---|---|---|---|---|---|---|---|---|---|
| | | | | distance to frontier | with the premium related to distance from the frontier | | | | |
| Babina, Fedyk, He & Hodson (2024) | Firm | United States, 2010–2018 | Firm-level AI investment inferred from the AI skills recorded in employee resumes; job postings | Sales growth; employment; market value; product innovation; concentration | AI-investing firms grow faster, primarily through product innovation, with growth concentrated among larger firms | Adoption | No | No | No |
| Czarnitzki, Fernández & Rammer (2023) | Firm | Germany, 2018 | Mannheim Innovation Panel, special module | Total factor productivity | Positive association between AI use and firm productivity | Adoption | No | No: four methods recorded, pooled into a breadth index | No: five application areas, pooled in index |
| Igna & Venturini (2023) | Firm | Europe, 1995–2016 | AI patent applications at the European Patent Office | AI inventive performance (patent productivity) | AI innovation shows strong dynamic returns and complementarities with earlier ICT knowledge | Innovation | Develops only (by construction: AI patents) | No: AI patents treated as one class | No |

## Appendix B. Firm-level data sources on AI and the four conceptual dimensions

| Data source | Examples of papers using this source | Novelty of AI application | Sourcing mode | Objectives of AI use | Type of AI technology |
|---|---|---|---|---|---|
| Official ICT surveys | Calvino & Fontanelli (2026); Fontanelli et al. (2025) | No | Yes | Inferred: from business functions | Yes |
| Innovation surveys (CIS) | Czarnitzki et al. (2023); Arenas Díaz et al. (2026) | Not of AI | No | Not of AI | No |
| Executive and business surveys | Acemoglu et al. (2022); Bonney et al. (2024); Hoffreumon et al. (2024) | No | In some surveys only | In some surveys only | In some surveys only |
| Worker surveys | Bick et al. (2026); Allen (2026) | No | No | No | In some surveys: generative only |
| AI patents | Igna & Venturini (2023); Giczy et al. (2022) | Yes (innovation) | Inferred: internally developed | Inferred: from claims | Yes |
| AI trademarks | Castaldi et al. (2026); Dernis et al. (2021) | Yes (new application) | No | Inferred: from trademark text | Inferred: from trademark text |
| Online job postings | Babina et al. (2024); Hunt et al. (2024) | No | Inferred: in-house hiring | No | Inferred: from named skills |
| Web-scraped firm data | Dahlke et al. (2024); Calvino et al. (2022) | No | No | Inferred: from website text | Inferred: from website text |
| Corporate disclosure text | Yu et al. (2026) | No | No | Inferred: from filing text | Inferred: from filing text |

Legend: *Yes*: the dimension can be explicitly measured. *Inferred*: the dimension is not explicitly measured, but it can partly be reconstructed from other information in the data source. *No*: the dimension can neither be explicitly measured nor inferred from the data.

## Appendix C. Thirty-six illustrative cases

This appendix points out one documented real case for each type of AI application of the combined typology presented in sections 4 and 5 (see table 7). As noted in section 5, the sources of these real cases are the following: the United States federal AI use case inventory, the United Kingdom Algorithmic Transparency Recording Standard, and the World Economic Forum's 2025 MINDS cohort. For each of the 36 types, this appendix states the organisation that implemented the AI application, what this does, and the source from which this case is taken.

A couple of notes on the coding are needed before the cases are presented. First, regarding the sourcing dimension, the sourcing of an AI application takes sometimes three forms rather than two: an organisation may purchase a working system from a supplier, it may assemble an application from commodity or open-source components, or it may develop the system itself. The typology in section 4 collapses these into two categories, because what matters for the economic effects of an application is whether the organisation holds the system and can alter how it behaves, or whether it uses a system that the supplier controls. An application is therefore placed in the internal development type when the organisation constructs and operates the system that delivers the application, and in the purchase type when the working system is supplied by a vendor and used as it comes, or when standard components are assembled to carry out a task rather than to build a system the organisation then runs. This binary distinction necessarily compresses some real-world cases. For example, open-source model weights are neither bought nor built in the ordinary sense; fine-tuning a commodity model lies between adaptation and development; and integrating a standard model into an organisation's own system is the most common configuration in all three inventories. In the classification of real cases presented below, where a case is on this boundary, the assignment is stated and justified in the description of the case itself (cases 17 and 23 are the closest pair in this respect).

Second, several of the records in the MINDS cohort name a pair of organisations, typically a technology provider and one client. As stated in section 2, a supplier and a user of the same system hold two distinct applications, which are classified independently. Each case below is therefore classified from the point of view of the organisation named at the beginning of its description, which in most paired records is the organisation that deploys the application.

### *Adoption*

**1. Adoption / Efficiency / Analytical.** In the US, the Veterans Health Administration deploys the Acumen Hypotension Prediction Index on the Edwards Lifesciences EV1000 clinical platform, which alerts clinicians to a patient's likelihood of a future hypotensive event during surgery. The system is purchased as a commercial medical device and used exactly as its manufacturer intends. The technology is predictive machine learning operating on physiological monitoring data; the objective is internal, improving the quality of clinical decisions rather than changing what the hospital offers. Nothing about the application is novel: the same device is sold to all hospitals.
(Source: US federal AI use case inventory, Department of Veterans Affairs)

**2. Adoption / Efficiency / Perceptual.** The same department uses GI Genius, a commercial computer-aided detection device that flags potential neoplastic lesions in real time during colonoscopy. The inventory record notes that multiple studies and meta-analyses have found that these devices increase adenoma detection rates, which is evidence that the application is established rather than novel. Purchased technology, perceptual AI technology, applied to an internal clinical process.
(Source: US federal AI use case inventory, Department of Veterans Affairs)

**3. Adoption / Efficiency / Generative.** In the US Department of the Interior, the Bureau of Trust Funds Administration adopted GitHub Copilot for software development and maintenance, explicitly to offset an increased workload following staff reductions. A commodity coding assistant, bought on subscription, applied to the standard task of writing and maintaining application code. This is the most common type of AI use of our typology.
(Source: US federal AI use case inventory, Department of the Interior)

**4. Adoption / Market / Analytical. In the UK,** HM Revenue and Customs operates Ask HMRC Online, a digital assistant that answers taxpayers' questions in natural language and escalates complex cases to human advisers, handling 5.48 million interactions in 2024 and 2025. The natural-language understanding model and the platform beneath it are supplied by Microsoft under procurement; the department builds only the conversational layer on top, which in our typology places the application in the purchase row. The technology is intent classification rather than generation. The objective is citizen-facing, i.e. providing a better service to taxpayers. The record describes the service as the first of its kind at the department, hence new to the organisation, not to the world, and therefore classified in the adoption cell.
(Source: UK Algorithmic Transparency Recording Standard, HM Revenue and Customs)

**5. Adoption / Market / Perceptual.** The GOV.UK One Login service verifies a user's identity remotely by matching a selfie against the photograph on an identity document and confirming that a live person is present, using convolutional neural networks and a dynamic liveness check that projects light patterns onto the face to avoid deepfakes. The capability is supplied entire by iProov under contract and integrated into the government application; roughly 100,000 identity journeys per month pass through it. Purchased perceptual technology, standard in the identity industry, directed at members of the public (citizens) rather than at internal process.
(Source: UK Algorithmic Transparency Recording Standard, Department for Science, Innovation and Technology)

**6. Adoption / Market / Generative.** In the US Social Security Administration, a question-and-answer bot on the agency's national telephone line answers callers' questions using approved material from the agency website, in English and Spanish, with calls handled by the bot, reducing waiting times for others. Purchased conversational technology, standard in design, directed outward to the public. It is the government analogue of a commercial customer-service chatbot.
(Source: US federal AI use case inventory, Social Security Administration)

**7. Adoption / Search / Analytical.** The US National Institutes of Health apply Pangolin lineage classification to SARS-CoV-2 genome sequences, processing a high volume of sequence data in real time to identify mutational patterns. Pangolin is a widely used public tool, recorded in the inventory as purchased from a vendor, and its application here is exactly what it was built for. The objective is research: the output feeds epidemiological understanding rather than an operational process. This places it in the search column despite the technology being routine.
(Source: US federal AI use case inventory, Department of Health and Human Services, National Institutes of Health)

**8. Adoption / Search / Perceptual.** The Geological Survey's Water Mission Area uses Google Cloud Vision to extract text from publicly available topographic maps as a callable function within an application. This is a commodity optical character recognition service, purchased through a cloud contract, that is applied to a digitisation task in support of research. Neither the technology nor the application is new to anyone, hence it is an adoption case.
(Source: US federal AI use case inventory, Department of the Interior)

**9. Adoption / Search / Generative.** The US Department of Energy procured Claude Enterprise to give research and technical staff access to large-context language models and code generation within a secured environment. An off-the-shelf enterprise assistant, deployed as sold, supporting scientific and engineering work. The search objective distinguishes it from the Copilot case above; the technology and the sourcing are similar.
(Source: US federal AI use case inventory, Department of Energy)

### *Co-invention*

**10. Co-invention / Efficiency / Analytical.** The company Fujitsu recently built a real-time optimisation capability for its own supply chain operations, reporting a 15 million dollar annual reduction in inventory cost, a 20 million dollar decrease in stock and faster assessment of disruptions. The underlying optimisation and forecasting methods are established; what was unprecedented was their integration into a global electronics supply chain at this scale and speed. Fujitsu is itself a technology producer, and the line between assembling and inventing is thinner for such a firm than for others; however, the case is placed by what the WEF report says it was done, i.e. the integration of established methods.
(Source: World Economic Forum, 2025 MINDS cohort)

**11. Co-invention / Efficiency / Perceptual.** PepsiCo deployed three-dimensional vision systems on edge machines in its manufacturing plants, achieving a 0.15% waste reduction in a high-precision context, and over 100 000 dollars of annual savings per production line. Machine vision for quality control is a mature technology; but applying it in this particular high-precision configuration was not standard practice in the sector. Bought technology, physical process, efficiency objective: hence, an illustration of perceptual co-invention.
(Source: World Economic Forum, 2025 MINDS cohort)

**12. Co-invention / Efficiency / Generative.** Social Medical Corporation Genshukai, a regional Japanese hospital group, worked with Fujitsu to place specialised AI agents across hospital management. The purpose was enforcing medical record standards, optimising bed allocation and coordinating with referring facilities. The reported effects are roughly 400 staff hours a month released, a 10% revenue increase and the elimination of compliance-driven fee refunds. The agent technology is Fujitsu’s, so the hospital assembles rather than invents; but agentic coordination of bed management and referral flow was not an available product, and the application had to be worked out for this setting. Bought technology, unprecedented application, internal objective.
(Source: World Economic Forum, 2025 MINDS cohort)

**13. Co-invention / Market / Analytical.** The National Institute of Clean and Low-Carbon Energy built a forecasting capability for electricity, carbon, coal and certificate markets, reporting a 50% reduction in price forecast error and 14 million dollars a year of additional revenue. The report describes what was built as "*an AI forecasting layer that pairs real-time data … with a language model specialized in energy policy and market analysis*". This counts as assembly rather than new model development. The application is nonetheless unprecedented: policy-aware forecasting across four coupled markets was not a product anyone could buy. The market objective rests on the report's own account of where the value lands: the system "*gives analysts and planners in energy markets earlier, clearer signals, driving increased sales revenue from trading optimization*". The forecast is used to sell more, or at a better moment, rather than to reduce costs. (Source: World Economic Forum, 2025 MINDS cohort)

**14. Co-invention / Market / Perceptual. In China,** Landing Med operates an AI cytology and telepathology system for population-based cervical cancer screening, reaching 91% coverage across Chinese provinces, correctly identifying positive cases at a rate 6% points above manual checking, and making pathology screening five times more productive. Computer vision on cytology slides is not a new technique; but delivering it as a population-scale screening service across a national territory was. The output is a service delivered outward, not internal efficiency. (Source: World Economic Forum, 2025 MINDS cohort)

**15. Co-invention / Market / Generative.** Ant Group operates a multimodal platform for clinical decision support and medical consultation, reporting over 90% auxiliary diagnostic accuracy, an 80% reduction in literature search time, and service to 160 million users, nearly one million doctors and over 5000 institutions. The models are assembled rather than invented, but a consumer-facing medical consultation service at this scale had no precedent.
(Source: World Economic Forum, 2025 MINDS cohort)

**16. Co-invention / Search / Analytical.** The University of California San Francisco Institute for Neurodegenerative Diseases, with SandboxAQ, built a discovery framework that virtually screened 5.6 million compounds in weeks and narrowed the pool to 7000 for laboratory testing, which is a 36-fold reduction in effort with hit rates 30 times higher than traditional screening. Virtual screening is an established method; assembling it into a framework that compresses a discovery pipeline by this factor was not. The output is downstream innovation.
(Source: World Economic Forum, 2025 MINDS cohort)

**17. Co-invention / Search / Perceptual.** Natural England built AI4Peat to map drainage features across England's upland peat (e.g. artificial channels, erosion gullies, exposed peat edges and restoration dams) from 12.5 cm aerial photography combined with LiDAR, using object detection for dams and semantic segmentation for the rest. The models are standard architectures taken from open-source libraries and trained on the agency's own imagery, which in our typology counts as assembling rather than building. What the agency contributed was the imagery and the mapping task rather than a system that it subsequently operates, which is why the case is assigned to the purchase category. The application, however, did not previously exist: the record states that before this work no national-scale map of peatland drainage features had been made. Assembled technology, unprecedented application, serving evidence rather than operations: the defining shape of co-invention in the search type of the typology.
(Source: UK Algorithmic Transparency Recording Standard, Natural England)

**18. Co-invention / Search / Generative.** Sanofi, with OAO, built an agent-based platform to accelerate the path from discovery to patient impact, reporting a 40% uplift in AI readiness, over 1400 employees activated and more than 1300 AI innovations generated, with a global significant sales increase. The technology is assembled from commercial components; using it as an engine for generating downstream innovation across a pharmaceutical organisation was novel. The objective is future innovation capacity rather than either cost or immediate revenue.
(Source: World Economic Forum, 2025 MINDS cohort)

### *Internalisation*

**19. Internalisation / Efficiency / Analytical.** The UK Department for Work and Pensions runs a supervised binary classification model that scores requests for Universal Credit advances for fraud risk, and it refers high-risk cases to human reviewers before payment. Supervised classification for fraud triage is thoroughly established; what the department has done is to apply this to its own claim archive. The record reports that the model has contributed to more than halving the scale of the problem.
(Source: UK Algorithmic Transparency Recording Standard, Department for Work and Pensions)

**20. Internalisation / Efficiency / Perceptual.** In the US, the Langley Research Center built a system called Autonomous Waiting Room Evaluation, which uses an existing security camera and an off-the-shelf object detection model to count people waiting at the Badge and Pass Office, and to alert staff when a threshold is exceeded. The detection model is not the agency's own, but the operating system that delivers the alerts was constructed and is run in-house, which is why we classify the application in the internal development type. Queue counting by camera is not a novel application, which is precisely why this belongs in the left column (new to the firm only) despite being built.
(Source: US federal AI use case inventory, NASA)

**21. Internalisation / Efficiency / Generative.** The US Securities and Exchange Commission built an internal retrieval-based assistant over documentation for its cloud environment, retrieving and summarising support material for technical staff. Retrieval-augmented generation over internal documentation is by now a commodity pattern; but the Commission built its own because the material and the environment are its own.
(Source: US federal AI use case inventory, Securities and Exchange Commission)

**22. Internalisation / Market / Analytical.** In the US Social Security Administration, Proactive Triage and Analysis of Hearings flags claims with a high likelihood of favourable outcome, and it refers them to human adjudicators for possible on-the-record decision, shortening the path for claimants. The record in the inventory lists better service to the public among the benefits, alongside case flow and consistency. The agency separately operates the Quick Disability Determinations model, which screens incoming applications for the same purpose, but was purchased from a vendor. The two make a good comparison: the same capability at the same agency, one bought and one built, but differing on the sourcing axis of the typology.
(Source: US federal AI use case inventory, Social Security Administration)

**23. Internalisation / Market / Perceptual.** US Citizenship and Immigration Services built a photo checking service into its online application system. When an applicant uploads an identity photograph with an application for employment authorisation, the service checks in real time whether the photograph meets the requirements of the card that will be printed from it, and tells the applicant if it does not. Services of this kind are sold commercially by several vendors; the capability is commercially available rather than new. But the agency assembled its own from open-source components, including face detection and object detection capabilities, trained in part on its own material. The agency constructed and operates the service itself, which is why the case is assigned to the internal development category even though the components are standard. The inventory record files it under service delivery and the output is addressed to the applicant, which is the basis for the citizen-facing placement (market).
(Source: US federal AI use case inventory, Department of Homeland Security)

**24. Internalisation / Market / Generative.** A system built in-house at the US Social Security Administration supports staff reviewing public comments by making preliminary designations as to whether a comment or attachment contains any of a set of predefined flagged content types (such as personally identifiable information, offensive content and other staff-defined categories) and it produces a catalogue with the quotations that informed each decision. The inventory record files it under service delivery, which is the basis for the citizen-facing placement in our typology. Automated triage of public comments is established practice across regulators; the agency built its own because the corpus and the categories are specific to it.
(Source: US federal AI use case inventory, Social Security Administration)

**25. Internalisation / Search / Analytical.** The US Centers for Disease Control' staff built classification methods to determine whether multiple bacterial isolates share the same antimicrobial-resistant plasmids, using genome sequencing data. Machine classification of sequence data is routine in bioinformatics; the implementation here is internal because the reference data and laboratory workflow are. The objective is research, and the application is standard within its field.
(Source: US federal AI use case inventory, Department of Health and Human Services)

**26. Internalisation / Search / Perceptual.** The staff of the Geological Survey in the US use deep learning to classify potential piping plover habitat along the Upper Missouri River, an assessment required annually under a Biological Opinion. The inventory record notes that the team has been using classification tools for this mapping over an extended period, which establishes it as settled practice rather than novelty. Built internally, perceptual technology type, in support of research and regulatory assessment.
(Source: US federal AI use case inventory, Department of the Interior)

**27. Internalisation / Search / Generative.** The US Agency for Toxic Substances and Disease Registry built a chatbot over its 180 toxicological profiles, allowing staff and enquirers to query chemical data, exposure pathways and health-assessment guidance directly rather than searching the documents by hand. Retrieval-augmented question answering over a document corpus is a commodity pattern; what is specific is the corpus, which is the agency's own and is why the system was built rather than bought. The inventory files it under science, and the output supports health assessment and enquiry response rather than a standard administrative process.

(Source: US federal AI use case inventory, Department of Health and Human Services, Centers for Disease Control and Prevention)

### *AI innovation*

**28. AI innovation / Efficiency / Analytical.** Hyundai, with DEEPX, developed ultra-efficient on-device AI semiconductors for autonomous robotics, allowing real-time inference on battery-powered delivery robots without cloud connectivity. The workload the chip is built to carry is perception and control for robotics; the objective is to reduce the cost and energy of running inference. The novelty is clear: custom silicon plus proprietary optimisation software represents a clear case of new technology development.
(Source: World Economic Forum, 2025 MINDS cohort)

**29. AI innovation / Efficiency / Perceptual.** Siemens, with EthonAI, developed an industrial AI framework for visual quality control, reporting over 10 million dollars of projected savings, a rollout across twenty factories, and inspection setup time reduced from days to minutes. The reduction of setup from days to minutes indicates method development rather than deployment of an existing product. In short, the case points to internal efficiency objective, perceptual AI technology, and frontier technology development.
(Source: World Economic Forum, 2025 MINDS cohort)

**30. AI innovation / Efficiency / Generative.** Cambridge Industries operates private large language model infrastructure with edge intelligence for urban maintenance and construction safety, reporting a 40% reduction in emergency road repair costs and roughly two million dollars a year saved per municipality. Two qualifications are important to classify this case. First, the system is not only generative: the report describes inspectors capturing road images that are analysed for surface damage and drones monitoring sites for safety risks, which is computer vision, with generation confined to translating safety manuals into site-specific guidance. In our classification, we assign a system that combines more than one technology type to the type of the component that carries the organisation's own contribution; here, that component is the privately operated generative infrastructure rather than the vision models applied to the imagery. Second, placing self-hosted model infrastructure in the development row rather than the assembly row is an assignment decision that we make because of the low-connectivity settings, where the engineering of the serving environment represents the substance of the work.
(Source: World Economic Forum, 2025 MINDS cohort)

**31. AI innovation / Market / Analytical.** TerraQuanta produced a meteorological forecasting system trained on 40 years of historical weather data combined with real-time observation, reported as 50,000 times more efficient than traditional models, and supplies it into energy trading markets. Horizon Power deploys it across a 2,000 gigawatt-hour trading portfolio. TerraQuanta built the model rather than assembling a layer over an existing one, which represents a case of new technology development. Its objective is to sell a new forecasting capability into a market that did not have anything of this kind.
(Source: World Economic Forum, 2025 MINDS cohort)

**32. AI innovation / Market / Perceptual.** The Ministry of Health of Saudi Arabia, with AmplifAI Health, developed thermal imaging analytics for diabetic foot risk assessment, reporting an 80% reduction in treatment costs and a 12-fold increase in screening capacity by enabling nurses to conduct AI-assisted screening while specialists concentrate on high-risk cases. Thermal imaging for this purpose has an existing research and commercial literature. But the WEF report records that the solution was validated in clinical trials and approved by regulators in Saudi Arabia, Bahrain and the United Arab Emirates, which is evidence of a new system built and taken through approval rather than procured. The objective is citizen-facing (i.e. market) in the sense adopted throughout for public organisations.
(Source: World Economic Forum, 2025 MINDS cohort)

**33. AI innovation / Market / Generative.** Tech Mahindra developed a sovereign multilingual framework for national language inclusion, raising conversational accuracy in Hindi and Bahasa from a 70% global benchmark to 92%, reducing translation errors by 35%, and supporting 3.8 million queries a month across citizen services, banking and healthcare. General-purpose models perform poorly in these languages, so models had to be built rather than adapted. The output is a service sold and deployed outward. This is an interesting illustration of AI innovation by a firm that is not an AI company.
(Source: World Economic Forum, 2025 MINDS cohort)

**34. AI innovation / Search / Analytical.** The firm Synopsys developed reinforcement learning and agentic workflows for semiconductor design, which the company AMD applied to its own chip development with designs reported as completed twice as fast, at five times lower cost and with fewer defects. The framework's unit is the application, so the two firms hold separate applications: the case placed here is Synopsys, since the WEF report states that AMD "*is* *harnessing Synopsys' reinforcement learning and agentic workflows*". Reinforcement learning applied to placement and routing at this level is new method development, and the objective is the generation of downstream innovation, with the recursive feature that the chips in question are those on which AI runs.
(Source: World Economic Forum, 2025 MINDS cohort)

**35. AI innovation / Search / Perceptual.** The closest case we find for this type in the three inventories is a Geological Survey project developing segmentation of land and water in time-series imagery to derive wave runup and total water level observations. The AI system is built internally and its purpose is research. For this case, though, the inventory source does not establish whether the segmentation models were developed or adapted from existing architectures; since adapting an existing model counts as assembling rather than building, the placement of this case in this position should be read with caution.
(Source: US federal AI use case inventory, Department of the Interior)

**36. AI innovation / Search / Generative.** At NASA, the Jet Propulsion Laboratory built ROSA, an agent that integrates with the Robot Operating System ecosystem to allow robots to be developed and operated through natural language, reading live telemetry and formulating responses from it. The agent architecture was developed internally and not procured; no equivalent existed; and the purpose is to accelerate the development of robotic systems. In short, this is an interesting case of a frontier generative application whose output is further engineering capability, built by a public research organisation rather than an AI company.
(Source: US federal AI use case inventory, NASA, with the project's public code repository)